\documentclass[%
      aps,
      prx, 
      reprint,
      twocolumn,
      showkeys,
      superscriptaddress]{revtex4-1}
      
\usepackage[utf8]{inputenc}
\usepackage[T1]{fontenc}
\usepackage{dcolumn}
\usepackage{bm}					
\usepackage{graphicx}
\usepackage{amsmath, amsfonts, amssymb, mathtools}	
\usepackage{latexsym} 				
\usepackage{xcolor}
\usepackage[caption = false]{subfig}
\usepackage[breaklinks, linkcolor = black]{hyperref}
\usepackage{tikz}
\usepackage{tabularx}
\usepackage{braket}
\usepackage{diagbox}
\usepackage[normalem]{ulem}
\usepackage{textcomp}
\usepackage{soul}
\usepackage{booktabs,eqparbox}

\begin{document}
\title{Charge quenching at defect states in  transition metal \\ dichalcogenide--graphene  van der Waals heterobilayers}

%
\author{Daniel \surname{Hernang\'{o}mez-P\'{e}rez}}

\email[]{daniel.hernangomez@weizmann.ac.il}

\affiliation{Department of Molecular Chemistry and Materials Science, Weizmann Institute of Science, Rehovot 7610001, Israel}

\author{Andrea Donarini}

\affiliation{Institute of Theoretical Physics, University of Regensburg, 93040 Regensburg, Germany}

\author{Sivan Refaely-Abramson}

\affiliation{Department of Molecular Chemistry and Materials Science, Weizmann Institute of Science, Rehovot 7610001, Israel}




\begin{abstract}
We study the dynamical properties of point-like defects, represented by monoatomic chalcogen vacancies, in WS\textsubscript{2}--graphene and MoS\textsubscript{2}--graphene heterobilayers. Employing a multidisciplinary approach based on the combination of \textit{ab initio}, model Hamiltonian and density matrix techniques, we propose a minimal interacting model that allows for the calculation of electronic transition times associated to population and depopulation of the vacancy by an additional electron. We obtain the ``coarse-grained'' semiclassical dynamics by means of a master equation approach and discuss the potential role of virtual charge fluctuations in the internal dynamics of impurity quasi-degenerate states. The interplay between the symmetry of the lattice and the spin degree of freedom through the spin-orbit interaction and its impact on charge quenching is studied in detail.
\end{abstract} 
 
\maketitle

\section{Introduction}\label{sec:intro}

In recent years, the study of van der Waals heterostructures \cite{Geim2013, Liu2016}, created upon stacking atomically-thin two-dimensional layers, %
has unveiled a rich variety  
of physical phenomena in these systems \cite{Jin2018}, ranging from sequential electron transfer through defects, studied using tunneling spectroscopy \cite{Steele2020, Steinberg2021}, to ultrafast interlayer charge transfer after photoexcitation with visible (yellow) light studied with time-resolved ARPES \cite{Yuan2018, Gierz2020, Gierz2021} or transient absorption spectroscopy \cite{Shuai2021}.
Such layered two-dimensional heterostructures can be grown
epitaxially in a controlled manner, revealing new families of materials that cannot be naturally found and whose properties extend far beyond the simple combination of the separated layers \cite{Jin2018}. 
Consequently, these systems offer a vast playground where the electronic and optical properties can be modified at the atomic scale due to combinations and variations in screening, electronic confinement or spin-orbit interaction.

Of particular interest are heterostructures of monolayer transition metal dichalcogenides (TMDCs) of the type XS\textsubscript{2} adsorbed on monolayer graphene. These systems can be considered as special type I heterostructures where one of the two layers is a semimetal, and therefore, both the conduction band minimum and the valence band maximum are located in the same layer and touch at individual high symmetry points in the Brillouin zone. Moreover, these heterostructures have the peculiarity of stacking graphene, a material characterized by an  extremely high carrier mobility \cite{Novoselov2005, Geim2008} and highly symmetric density of states (DoS) close to its charge neutrality point, with a strongly confined monolayer semiconductor with direct band gap and which can have a sizable spin-orbit splitting \cite{Lanzara2015}.
%

Strong confinement in two-dimensional heterostructures makes them particularly sensitive to the potential created by impurities or vacancies (defects). These vacancies are known to produce in-gap states with clear fingerprints that can be measured by local probes \cite{RefaelyAbramson2018} as well as affect the heterostructure optical properties. These optical properties are also well known to be related to the large pristine \cite{Urbaszek2018, Goryca2019} and defect-induced \cite{RefaelyAbramson2018} exciton binding energies and show valley circular dichroism due to valley-spin coupling by means of spin-orbit interaction \cite{RefaelyAbramson2019, Mitterreiter2021, Cao2012, Glazov2021}. 

From the electronic point of view, defects created by substitutions or vacancies can be considered as ``electron traps''. They have been anticipated to decrease the electronic conductivity and electronic mobility when the Fermi energy is close to the localized defect levels, hybridized, for example, with the continuum from the band extrema \cite{Bassani1974, Lischner2020}. 
For similar reasons, defect scattering channels have also been hypothesized to be the source of longer lifetimes of electronic states compared to hole ones in WS\textsubscript{2}--graphene interfaces after ultrafast charge separation induced by photoexcitation \cite{Gierz2020, Gierz2021, Wang2021} as they are expected to efficiently pin down the photogenerated charge carriers. 
However, as was recently shown by some of us for the case of oxide surfaces \cite{Steinitz2022}, the role of defects may be subtle and cannot be always considered as trivial charge traps simply increasing localization or decreasing photoconductivity.

%
%

Defects created by vacancies in the regular lattice of TMDCs can be seen also as quantum dots \cite{Reimann2002} with the symmetry provided by the host lattice determining their shell structure. In these ``artificial atoms'', the discrete electronic charge as well as the interplay between Coulomb interaction, spin-orbit coupling and quantum statistics manifests in several well-studied physical effects (statistical crossover in the many-body interference blockade regime \cite{Groth2006}, channel blockade \cite{Haug2016}, spin blockade \cite{Platero2010}, geometrical charge frustration \cite{Mahalu2013} or triplet-singlet transitions induced by superexchange \cite{Korkusinski2007} among many others) and applications (spin qubits \cite{Gaudreau2012} or current rectifiers \cite{Stopa2002}). %
In particular, many-body quantum interference has been the subject of recent research \cite{Donarini2009, Donarini2010, Donarini2019} as well as its connection with the quantum dot symmetry and symmetry-induced degeneracies.
Such type of dots  can be studied both experimentally and theoretically for lateral heterostructures \cite{Haug2016, Gaudreau2006, Mahalu2013, Pan2012, Niklas2017}.
Vacancies in TMDCs and TMDC heterostructures thus open the possibility of studying the electronic and spin properties of these systems in the atomistic limit.  

In this paper, we study the charge transfer processes occurring in XS\textsubscript{2}--graphene (Gr) heterobilayers (here, X = Mo, W) with chalcogen vacancies. These vacancies, understood in the dilute limit, act as isolated defects or ``dots'' with the symmetry provided by the hosting lattice. The vacancy states are tunnel-coupled to a graphene reservoir which provides a structured DoS. We compute the electronic transition times for charging and discharging of the empty levels of the vacancy, due to tunneling from and into the graphene layer, with a combination of \textit{ab initio}, model Hamiltonian and density matrix-based methods. The transition rates for second order processes in the tunneling Hamiltonian are employed to understand the semiclassical many-body dynamics. We also discuss the role of degeneracies in the potentially coherent dynamics at the dot.
Finally, we examine the interplay between orbital (lattice ``symmetry'') and spin degrees of freedom that can occur by means of spin-orbit interaction. The population dynamics for the non-trivial charge transfer mechanism and its consequences for the semiclassical quantum dot dynamics are analyzed in detail. Overall, we show how the combination of different theoretical techniques can be employed to study quantum dot dynamics at complex interfaces, with realistic physical parameters reflecting the specific system structure.
This work will also serve as a basis to understand more subtle aspects of the quantum dot dynamics and as a guidance to the interpretation of experimental observations.

The paper is organized as follows: in Sec. \ref{sec:dft}, we present density functional theory (DFT) band structure calculations for the  MoS\textsubscript{2}--Gr and WS\textsubscript{2}--Gr heterostructures upon which our model will be constructed. In Sec. \ref{sec:model} we discuss the low energy model Hamiltonian used to analyze our DFT results.  A summary of the kinetic theory used in this manuscript is briefly presented in Sec. \ref{sec:transport}. The results for the dynamics and the electronic transition rates are shown in Sec. \ref{sec:results}, considering separately the case of weak and strong spin-orbit interaction. We conclude this paper in Sec. \ref{sec:conclusion} and devote the Supplemental Material (SM) to computational details, explicit analytical derivations and to supplementary results.


\section{Density Functional Theory Band Structure Analysis}\label{sec:dft}

\subsection{Geometry}
%

\begin{figure}
    \includegraphics[width=1.0\linewidth]{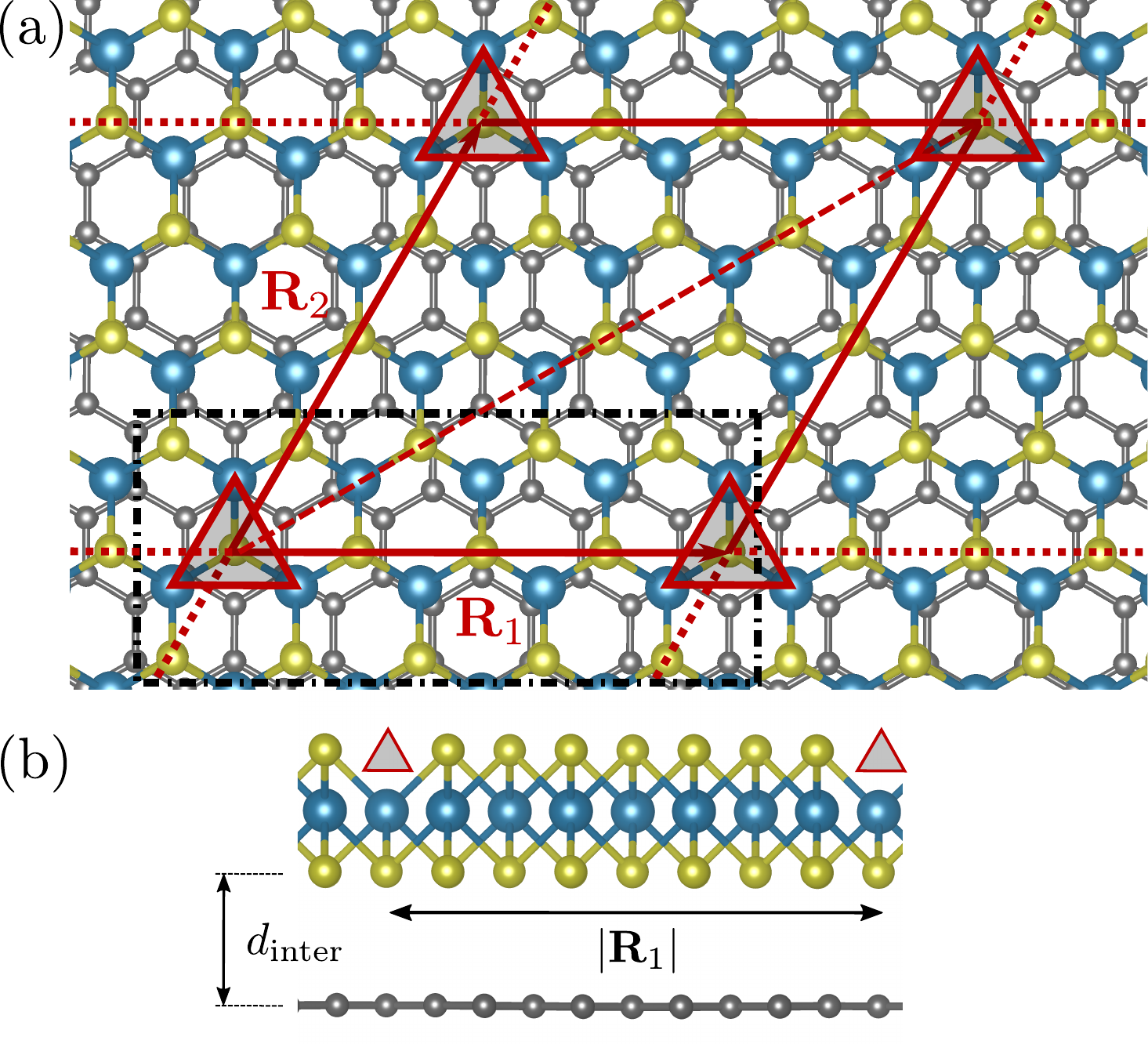}      
    \caption{(a) Top view of the XS\textsubscript{2}--Gr supercell (here X = W). The $4\times 4/ 5\times 5$ supercell is indicated by the red solid parallelogram, the position of the single chalcogen vacancy (and its symmetry) is marked by the red triangle located at the origin of the unit cell, appearing four times due to the supercell periodicity in this panel. The dotted lines represent the borders of adjacent supercells.
    The dashed red line along the main diagonal indicates that graphene adsorption on the TMDC reduces the global symmetry of the system from C\textsubscript{3v} to C\textsubscript{s}, \textit{i.e.} the  only symmetry element is the reflection plane perpendicular to the heterostructure and bisecting the angle between the supercell basis vectors. 
    %
    (b) Side view of the WS\textsubscript{2}--Gr heterostructure cut along the dashed-dotted black box marked in panel (a).
    The missing atom corresponds to the chalcogen vacancy in the supercell, represented here by a red triangle.
    }\label{f1} 
\end{figure}

To study the role of isolated defects in XS\textsubscript{2}--Gr heterobilayers, we adopt the supercell approach.
The supercell employed in this work is shown in Fig. \ref{f1} (a).
This supercell is made from a heterobilayer formed by a TMDC of the type XS\textsubscript{2} (where X = W, Mo) and graphene.
The commensuration of the corresponding lattices in each layer requires 
to have $4\times4$ elementary cells for XS\textsubscript{2} and $5\times5$ 
for  graphene in the supercell.
The supercell was previously optimized using DFT, see details in Methods section.
The in-plane lattice constant for the XS\textsubscript{2} monolayer is equal to $3.15$ \AA, while graphene has an in-plane lattice constant of $2.52$ \AA\, (the distance between nearest neighbor carbon atoms of $1.455$ \AA).
In other words, the graphene lattice is strained by $\sim 2.4$\% with respect to an isolated relaxed graphene monolayer while we keep TMDC layer almost not strained with respect to the experimental lattice constant \cite{Schutte1987}. In this sense, we note the difference from other calculations performed in previous works with the same type of commensurate supercell \cite{Yuan2018}, in which the strain is distributed between the two layers of the TMDC--Gr heterostructure 
with similar lattice mismatch.
The supercell lattice vectors therefore have a length of $|\mathbf{R}_1| =  |\mathbf{R}_2| \simeq 12.6$ \AA \,.
The optimized interlayer distance between the XS\textsubscript{2} and the graphene layers in the supercell
is $3.44$ \AA. In order to prevent spurious interaction between the periodic replicas, an out-of-plane vacuum layer of width of $\sim 17$ \AA\, is employed. 
We consider a single chalcogen vacancy per unit cell (this would correspond to a $\sim 3\%$ vacancy concentration). 
The vacancy is located on the opposite side, in correspondence to a hollow position of the graphene layer
and, for convenience, it is chosen to be the origin of the supercell.
In our calculations, the supercell with a single vacancy has been created by removing a chalcogen atom from the previously optimized non-defected
periodic supercell.
We have checked that structural relaxation of the heterostructure lattice in the presence of  the vacancy does not change the symmetry properties, only shrinking the defect size by roughly $\sim 5\%$ due to a reduced metal-metal (X--X) distance close to the vacant sulphur atom while preserving qualitatively the shape and nature of the DFT band structure.  %

\subsection{Band structure without spin-orbit interaction}

\begin{figure*}
    \centering
    \includegraphics[width=0.8\linewidth]{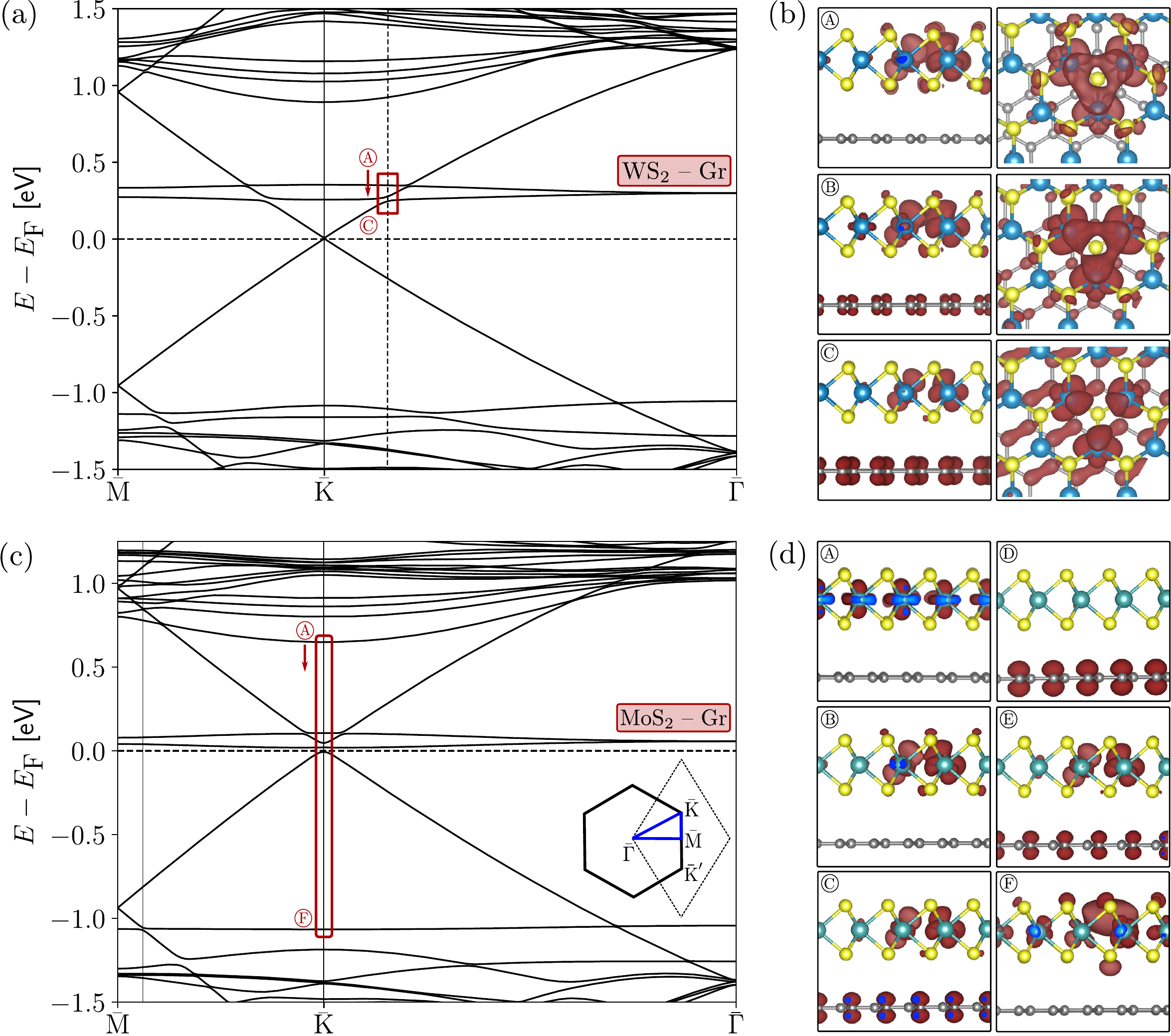}
    \caption{(a) Band structure of the WS\textsubscript{2}--Gr heterostructure in the absence of spin-orbit interaction computed along the $\bar{\textnormal{M}}-\bar{\textnormal{K}}-\bar{\Gamma}$ path in the supercell. The chalcogen vacancy creates three defect bands, two of them appearing in the pristine WS\textsubscript{2} gap and \textcolor{black}{hybridizing with} the graphene Dirac cone centered at $\bar{\textnormal{K}}$ and setting the charge neutrality point.
    (b) Kohn-Sham pseudo-charge density ($|\Psi|^2$) for the empty states in the red box in panel (a), labeled by {\large \textcircled{\small A}}, {\large \textcircled{\small B}} and {\large \textcircled{\small C}} from highest to lowest energy. The left column shows the side view, while the right column displays the top view of the corresponding Kohn-Sham density.
    (c) Same as in (a) but for the MoS\textsubscript{2}--Gr heterostructure. The inset shows the path in the hexagonal and rhombohedrical mini-Brillouin zone. As in the WS\textsubscript{2}--Gr interface, the chalcogen vacancy creates three defect bands, two of them appearing in the MoS\textsubscript{2} pristine gap close to the Dirac point and opening a gap at the charge neutrality point of graphene located at $\bar{\textnormal{K}}$.
    (d) Same as in (b) but for the states in the red box in panel (c), labeled from {\large \textcircled{\small A}} to {\large \textcircled{\small F}} from highest to lowest energy.
    }\label{f2db}
\end{figure*}

We start by considering the DFT band structure of the periodic supercell in the absence of spin-orbit interaction.
The computational details of our \textit{ab initio} calculations are given in Sec. I of the SM.
The band structure of the WS\textsubscript{2}--Gr interface is shown in Fig. \ref{f2db} (a). 
We clearly observe the superposition
of the band structures of graphene (with its well-known Dirac cone located
at the $\bar{\textnormal{K}}$ point of the mini-Brillouin zone) and WS\textsubscript{2} monolayers, similarly to previously reported calculations \cite{Yuan2018}. 
For our supercell, the Dirac cone that sets the Fermi energy of the heterobilayer (and corresponds to the graphene charge neutrality point) is located in the WS\textsubscript{2} pristine bandgap. Note that since the crystal structure of each individual layer has an hexagonal unit cell, the Brillouin zone is also hexagonal and matching at the edge for this supercell configuration sets the point $\textnormal{K}$ of the unfolded Brillouin zone to always map into $\bar{\textnormal{K}}$ of the mini-Brillouin zone by construction \cite{Naimer2021} (other high symmetry points may also map onto $\bar{\textnormal{K}}$). 
The non-defect conduction 
and valence bands of WS\textsubscript{2} are located at $\sim 1$ eV
higher and lower in energy, respectively, from the graphene $\bar{\textnormal{K}}$ point.
The sulphur vacancies generate three defect bands stemming from the bond orbitals of the neighbouring metal atoms \cite{Noh2014, Robertson2013, Idrobo2013}. In addition, as a consequence of the C$_{3v}$ symmetry of the sulphur vacancy, the defect bands comprise two empty and quasi-degenerate bands in the pristine WS\textsubscript{2} gap and one occupied band located in the valence region (see Sec. II of the SM for a general discussion on elementary symmetry properties).
At the $\bar{\Gamma}$ point, the two empty in-gap states are located at $\sim 0.3$ eV above the graphene charge neutrality point. Here, the slightly broken symmetry in the supercell (see Fig.\ref{f1}) manifest itself in a splitting of $\sim 1$ meV between the energy levels.
The vacancy states are mostly composed of $d$-orbitals of the W atoms closer to the vacant S atom. This situation was previously shown in other TMDCs such as MoS\textsubscript{2} (see Refs. \onlinecite{Noh2014, Robertson2013, Idrobo2013} and below).
While the in-gap defect bands are relatively flat due to the spatial localization of the defect states in real space, they are still weakly dispersive close to the Dirac cone. This results from residual defect-defect interaction between neighbouring supercells when the TMDC layer has $4\times4$ XS\textsubscript{2} unit cells, see Supplemental Material in Ref. \onlinecite{RefaelyAbramson2018}.
We note that the in-gap defect bands hybridize with the graphene Dirac cone close to the $\bar{\textnormal{K}}$ point, generating
anticrossings with characteristic level splittings.  The different size of the two anticrossings is attributed to the symmetry
properties of each of the states, being even and odd with respect to \textcolor{black}{a reflection  on a plane perpendicular to the heterostructure and located along the supercell main diagonal, see Fig. \ref{f1}.}
\textcolor{black}{We have checked that the electronic band structure is robust to the position of the vacancy, \textit{i.e.} the band structure to be qualitatively invariant and the anti-crossing position and size are well-preserved, as globally breaking spatial symmetries in the supercell due to a change of the position of the vacancy does not substantially alter the local symmetry properties of the associated in-gap localized states.}

In  Fig. \ref{f2db} (b), we show the (pseudo)-densities of the Kohn-Sham states, $|\Psi|^2$, marked by the red box in Fig. \ref{f2db} (a) for the corresponding \textbf{k}-point (dashed vertical line). 
As anticipated above, the hybridization between the two layers occurs at specific \textbf{k}-points in the supercell Brillouin zone (forming an avoided ring around the graphene Dirac point due to the almost perfect conical shape of the graphene bands). This hybridization yields in-gap states with weight on both layers of the heterostructure at these avoided crossings [see panels {\large \textcircled{\small B}}, {\large \textcircled{\small C}} of Fig. \ref{f2db} (b) in comparison to panel {\large \textcircled{\small A}}]. 
We note that hybridization can also occur far from the Fermi level, for example, in the valence band region close to the $\bar{\textnormal{M}}$ point, where coupling occurs between extended states on the WS\textsubscript{2} and  graphene layers as can be anticipated from the local energy level splitting. 


The scenario described for the WS\textsubscript{2}--Gr interface is similar to the one found in the MoS\textsubscript{2}--Gr heterostructure, see Fig. \ref{f2db}, panels (c) and (d). 
In Fig. \ref{f2db} (c), we display its DFT band structure. The geometry employed was obtained from the WS\textsubscript{2}--Gr supercell by substitution of the W atoms by Mo and keeping the same lattice parameters (the experimental lattice parameters for MoS\textsubscript{2} and WS\textsubscript{2} are almost the same, being 3.15 \AA\, for the former \cite{Nicklow1975} and 3.153 \AA\, for the latter \cite{Schutte1987}).
We observe that, similar to the 
WS\textsubscript{2}--Gr interface, the local C$_{3v}$ symmetry of the vacancy yields the anticipated three defect 
bands, two of them being located in the pristine MoS\textsubscript{2} bandgap (about $50$ meV above the charge neutrality point for $\bar{\Gamma}$) and one in the valence band region \cite{Noh2014, Robertson2013, Idrobo2013,RefaelyAbramson2018}. 
Interestingly, the graphene Dirac cone is now gapped (by about $50$ meV) due to the hybridization of graphene and MoS\textsubscript{2} defect bands occurring very close to the charge neutrality point. 
In Fig. \ref{f2db} (d), we display the corresponding Kohn-Sham pseudodensities, $|\Psi|^2$,  for the states located in the red box in panel (c). 
As expected from the presence of anti-crossings in the energy spectrum, we find again interfacial hybridized states with substantial wavefunction weight on
both layers of the heterostructure, see panels {\large \textcircled{\small B}} and {\large \textcircled{\small D}}.

%

\subsection{Band structure with spin-orbit interaction}\label{sec:bands_SOC}

\begin{figure}
   \includegraphics[width=1.0\linewidth]{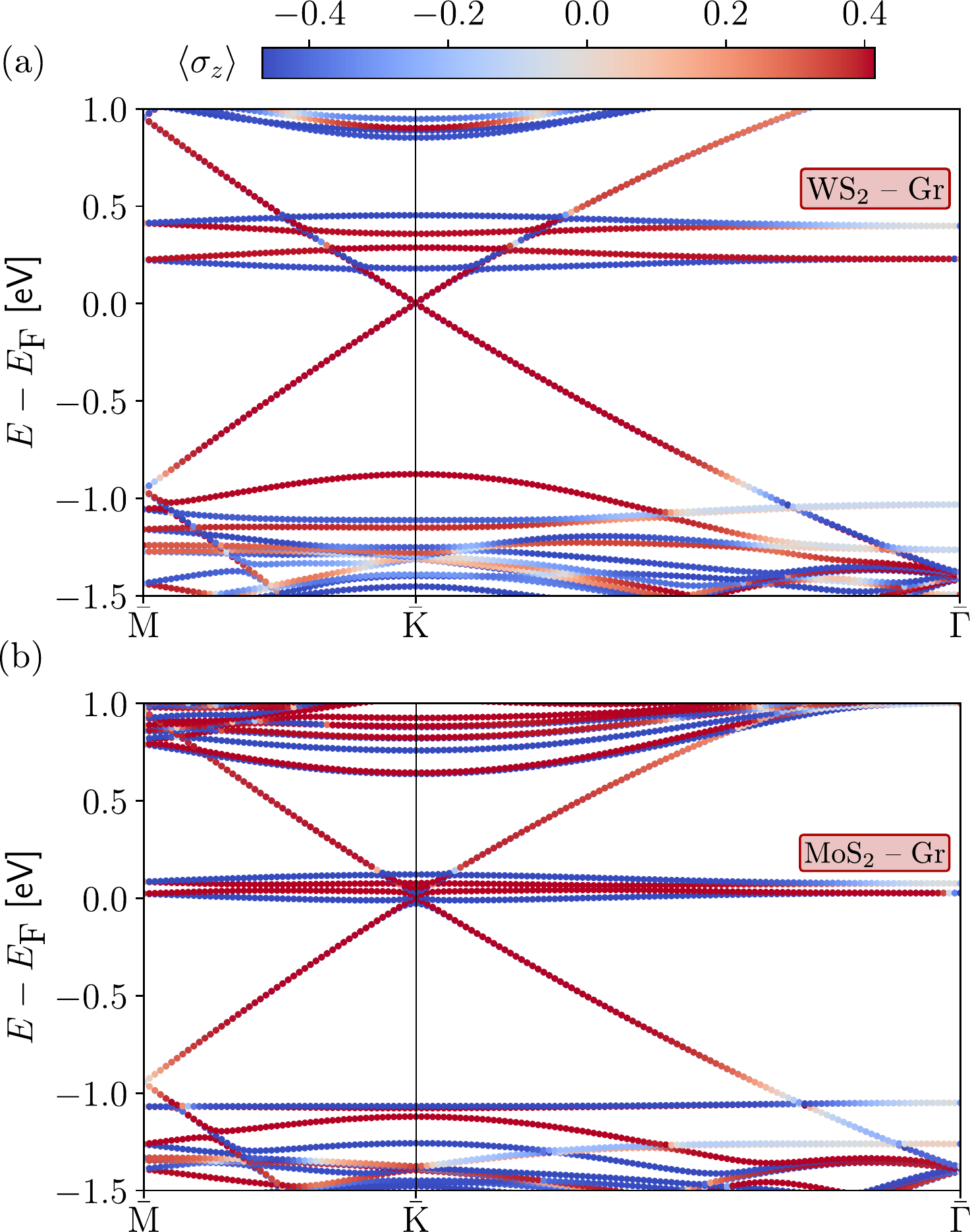}      
    \caption{Band structures of the XS\textsubscript{2}--Gr heterostructures in the presence of spin-orbit interaction computed along the $\bar{\textnormal{M}}-\bar{\textnormal{K}}-\bar{\Gamma}$ path in the supercell. Panel (a) corresponds to the case X = W, panel (b) to the case X = Mo. The spin-orbit splitting of the defect energy levels yields two degenerated levels at the $\bar{\Gamma}$ point, which later splits into four levels clearly visible closer to the graphene Dirac cone centered at $\bar{\textnormal{K}}$. 
    The color code represents the spin magnetization projected in the direction perpendicular to the heterobilayer.
    The spin-orbit split defect levels have a well-defined spin direction in the vicinity of the $\bar{\textnormal{K}}$ point and become spin-degenerated both at high symmetry points $\bar{\textnormal{M}}$ and $\bar{\Gamma}$. The graphene Dirac cone remains spin-degenerated in the whole Brillouin zone as expected due to the small spin-orbit coupling present in the carbon atom as well as its regular structure.
    }\label{f6} 
\end{figure}

In the presence of spin-orbit interaction, the valence and conduction bands split and the DFT band-gap
from pristine WS\textsubscript{2} is slightly reduced to $\sim 1.75$ eV from $\sim 1.95$ eV, see Fig. \ref{f6} (a).
For comparison, the calculated WS\textsubscript{2} bandgap (computed at the G\textsubscript{0}W\textsubscript{0} level for a $5\times 5$ supercell without adsorbed graphene) is equal to $2.8$ eV while the experimental bandgap equals to $2.5$ eV \cite{RefaelyAbramson2018}. In general, not only screening but adsorption by graphene is expected to renormalize the G\textsubscript{0}W\textsubscript{0} band gap as well \cite{Jain2018}. 
Therefore, in our calculations we underestimate the bandgap due to the well-known approximations in the exchange-correlation part of the DFT functional by at most $\sim 0.75$ eV.
The spin-orbit interaction also splits the pair of in-gap defect levels by
0.15 eV at the $\bar{\Gamma}$ and $\bar{\textnormal{M}}$  points, suggesting that the spin-orbit coupling is strong. 
This spin-orbit splitting is well understood from the combination of the action of  time-reversal and translational symmetry \cite{Zhu2011}.
In particular, note that at the time-reversal invariant points for the honeycomb lattice $\bar{\Gamma}$ and $\bar{\textnormal{M}}$, all bands need to be doubly degenerated \cite{Zhu2011, Fu2007},
%
while close to the spin-degenerated Dirac cone, the vacancy in-gap states split due to their small but finite dispersion yielding a spectral structure of four spin-orbit split energy levels.  

The scenario is analogous for 
the MoS\textsubscript{2}--Gr interface, as displayed in Fig. \ref{f6} (b). Here, however, the spin-orbit coupling reduces the pristine band gap to  $\sim 1.7$ eV from $\sim 1.85$ eV as weaker spin-orbit interaction induces smaller splitting of the conduction and valence bands of MoS\textsubscript{2} compared to WS\textsubscript{2}.
A value of $\sim 2.8$ eV has been reported for the isolated MoS\textsubscript{2} monolayer at the G\textsubscript{0}W\textsubscript{0} level \cite{Qiu2013}. In the case of a non-defected interface with a similar supercell size than the one employed here, previous calculations in the literature yield a pristine band gap upon adsorption of $1.73$ eV at the DFT level and $2.43$ eV at the G\textsubscript{0}W\textsubscript{0} level \cite{Thygesen2015}. Therefore, we expect an underestimation of the band gap by $\sim 0.7$ eV.
The quasi-degenerated defect bands are split by $\sim 0.05$ eV at the $\bar{\Gamma}$ and $\bar{\textnormal{M}}$  points, roughly three time smaller than in  WS\textsubscript{2} due to weaker spin-orbit interaction
in  MoS\textsubscript{2} compared to WS\textsubscript{2}.
For the same reason, the occupied defect band barely lifts its spin-degeneracy and the spin-orbit splitting is $\sim 10$ meV at the $\bar{\textnormal{K}}$ point. 
Moreover, due to spin-orbit coupling, the lower empty defect band now appears below the graphene charge neutrality point. 
In other words, the system
acquires partially a metallic character. However, note that this occurs only close to the $\bar{\textnormal{K}}$ point, where the defect bands bend due to residual defect-defect interaction between the periodic supercells, and far from the $\bar{\Gamma}$ point, which is the point of interest for the defect states in the isolated vacancy limit. Note that this is a non-standard usage of the supercell approach, as we are employing the supercell (which contains vacancies in a $\sim 3$\% concentration) in order to study the dynamics at the level of the isolated vacancy.
We also note that among the small effects of the geometry optimization in the presence of the vacancy is to revert the situation of partial occupancy of the vacancy energy bands by shifting the defect states again above the Dirac $\bar{\textnormal{K}}$ point. Because this effect is not relevant for the physics discussed in the rest of the manuscript, in our model we shall simply consider always two empty in-gap defect levels ({\it i.e} four states, once spin is taken into account).

To understand the impact of spin-orbit coupling beyond the energy spectrum, we also computed the spin magnetization of the defected XS\textsubscript{2}--Gr heterostructures. The results for the spin expectation value in the direction perpendicular to the heterobilayer are shown by the color code of the electronic band structure in Fig. \ref{f6}.
An important observation that results from this calculation and that will be employed later for the modelling of the dynamics, is that hybridization of the vacancy levels and graphene preserves the spin orientation. This can be easily seen in  Fig. \ref{f6} where the spin up defect level couples only to the spin up component of the graphene Dirac cone (here represented in red color); similarly, the spin down defect level only hybridizes with the spin down component (here shown in blue color). Therefore, we anticipate a competition between the spin and ``orbital'' (the angular momentum given by the lattice symmetry) degree of freedom to be manifested in the dynamical features of the population and depopulation of the vacancy once strong spin-orbit interaction is taken into consideration.

\section{Model}\label{sec:model}

In order to microscopically describe the  dynamics associated with the vacancy embedded in the TMDC--Gr interface, we consider an \textit{ab initio} inspired model 
for a quantum dot coupled to a graphene ``lead'' (\textit{i.e.} reservoir or bath) in a system-bath setup. 
Our model Hamiltonian is derived using electronic structure information, namely, 
 the hybridization matrix elements between the TMDC and graphene monolayers are calculated by means of the combination of DFT and symmetry analysis, with the 
 position of the vacancy levels being derived from DFT, which could potentially be corrected by a GW calculation. 

We thus follow the standard partitioning of  transport  systems, in which the  Hamiltonian is composed of three parts, 
\begin{equation}\label{eq:hamiltonian_sys}
\hat{H} = \hat{H}_\textnormal{sys} + \hat{H}_\textnormal{lead} + \hat{H}_\textnormal{tun}.
\end{equation}
Here, the system Hamiltonian corresponds to our vacancy (``dot'') described as 
\begin{equation}
 \hat{H}_\textnormal{sys} = \sum_{i\sigma}  \epsilon_{i\sigma} \hat{d}^\dagger_{{i}\sigma} \hat{d}_{i\sigma} + \dfrac{E_c}{2} \hat{N}(\hat{N}-1) +  \hat{H}_\textnormal{SOC}.
\end{equation}
The first term describes an effective single-particle vacancy Hamiltonian without spin-orbit interaction.
We represent by $\hat{d}^\dagger$ (resp. $\hat{d}$) the creation (resp. annihilation) operator of an electron in the dot, labeled by the orbital index $i$ and spin $\sigma$; $\epsilon_{i\sigma}$ are the corresponding single-particle effective energies of the vacancy assumed to be independent of the quasi-momenta associated to the supercell \footnote{As discussed above and in previous work, we neglect, for the modelization, the spurious defect-defect interaction between vacancies belonging to different cells and thus we assume non-dispersive bands $\epsilon_{i\mathbf{q}\sigma} \simeq \epsilon_{i\sigma}$ as the dispersion comes from spurious defect-defect interaction between vacancies that belong to different supercells.}.
The second term corresponds to the Coulomb interaction at the dot within the constant interaction picture. Here, $E_c$ is the so-called charging energy and $\hat{N} = \hat{d}^\dagger_{i\sigma}\hat{d}_{i\sigma}$ the total number operator. 
By construction, the charging energy %
participates in the physical processes only if more than one electron populates the vacancy. Due to the size of the defect impurity (see Fig. \ref{f2db}) we focus on the limit in which $E_c$ is the largest energy scale of the system.
Finally, the spin-orbit coupling term accounts for the splitting observed in the \textit{ab initio} calculations and its form and impact on the dynamics will be discussed in Sec. \ref{sec:soc}. 

The ``lead'' Hamiltonian corresponds to the graphene layer that acts as an fermionic reservoir/bath with
\begin{equation}
 \hat{H}_\textnormal{lead} = \sum_\mathbf{k\sigma} \epsilon_\mathbf{k\sigma} \hat{c}^\dagger_\mathbf{k\sigma} \hat{c}_\mathbf{k\sigma}.
\end{equation}
Here, $\hat{c}^\dagger$ (resp. $\hat{c}$) corresponds to the 
electronic creation (resp. annihilation) operators in the graphene layer and $\epsilon_\mathbf{k\sigma}$ is the low-energy graphene energy dispersion relation. Close  to the $\bar{\textnormal{K}}$ point, the graphene dispersion relation is isotropic and linear, \textit{i.e.} $\epsilon_{\mathbf{k}\sigma} = \hbar v_\textnormal{F} |\mathbf{k}|$,  and with slope proportional to the Fermi velocity, $v_\textnormal{F}$. 

Finally, the tunneling Hamiltonian, that combines the bath and system operators, is given by
\begin{equation}\label{eq:tunneling}
    \hat{H}_{\rm tun} = \sum_p \sum_{i\sigma} \sum_{\mathbf{k}} p  t^p_{i\mathbf{k}\sigma} \hat{d}^p_{i\sigma} \hat{c}^{\bar{p}}_{\mathbf{k}\sigma}, 
\end{equation}
where $p = \pm$ is employed as a short-hand notation to label the creation ($+$) and annihilation ($-$) operators, together with the conventions $\hat{c}^+ = \hat{c}^\dagger$, $\hat{c}^{-} = \hat{c}$. We also adopt the notation $\bar{p} = -p$. 
The tunneling Hamiltonian describes the coupling between the vacancy at the TMDC layer and the graphene reservoir through the set of hybridization (or tunneling) matrix elements, $t^{\pm}_{i\mathbf{k}\sigma}$. These matrix elements are interpreted as the coupling between a state with momentum $\mathbf{k}$ and spin $\sigma$ in the graphene layer and a defect state in the XS\textsubscript{2} layer with quantum numbers $i$ and $\sigma$.
By definition, the tunneling amplitudes satisfy $t^+ = (t^{-})^\ast$. Note that tunneling by definition preserves the spin orientation due to the small vacuum distance that separates the two layers of the heterostructure as it was seen in Sec. \ref{sec:bands_SOC} and further discussed in Sec. \ref{sec:results}. %

So far we have only considered  DFT information in our model. However, a more realistic parametrization would include the correction to the position of the defect levels as obtained from a G\textsubscript{0}W\textsubscript{0} calculation. Under the assumption that the hybridization matrix elements are not affected at this level, as they essentially depend on the defect and graphene wavefunctions, it is possible to estimate the impact of screening in the many-body transition  rates by shifting $\epsilon_{i\sigma} \rightarrow \epsilon_{i\sigma} + \delta E$, where $\delta E$ would approximately be half of the change from the G\textsubscript{0}W\textsubscript{0} to the DFT pristine bandgap. 
In in Sec. IV of the SM we will discuss, by means of an illustrative case, how the model depends on the position of the defect energy levels. Reciprocally, we could also use the levels shift of a DFT band structure as a free parameter, if fit to experimental data is required, therefore, providing guidance in the interpretation of experimental results. 

\section{Kinetic Equation}
\label{sec:transport}

\subsection{Generalities}
We investigate the charge quenching  of the strongly interacting impurity-bath system by means of the generalized master equation for the reduced density operator \cite{Braun2004, Schon2005, Begemann2008, Wacker2011, Donarini2019, Donarini2021}. This approach allows for the description of interference effects even for strongly interacting systems (see Ref. \cite{Donarini2021} and references therein). Moreover, its recent application to THz-STM \cite{Donarini2021b} suggests a natural extension of the present study of XS\textsubscript2 impurities to setups with time dependent driving. Here, we briefly summarize the main steps to the derivation of a kinetic equation for the reduced density operator.

We start from the Liouville-von Neumann equation \cite{Sakurai2021} for the density matrix, which can be written in a compact form as
$
 \dot{\hat{\rho}} = \mathcal{L} \hat{\rho}
$, with $\hat{\rho}$ being the density operator and the dot the time derivative. The object $\mathcal{L}$ is the so-called Liouvillian superoperator defined as
\begin{equation}\label{eq:liouvillian}
 \mathcal{L} := -\dfrac{i}{\hbar}[\hat{H}, \bullet],
\end{equation}
with $[\bullet, \bullet]$ the commutator and $\hat{H}$ the total Hamiltonian combining the system, bath and hybridization terms, see Eq.  \eqref{eq:hamiltonian_sys}.
The equation of motion for the reduced density matrix, which contains degrees of freedom of the system only, can be obtained after tracing out the degrees of freedom of the fermionic bath, $\hat{\rho}_\textnormal{red} = \textnormal{Tr}_\textnormal{lead} \hat{\rho}$. 
The bath is assumed to be at local thermodynamical equilibrium, characterized by a temperature $T$ and chemical potential $\mu$, and, therefore, it is described by the equilibrium density matrix
\begin{equation}
 \hat{\rho}_\textnormal{lead} = \dfrac{{\rm e}^{-\beta(\hat{H}_\textnormal{lead} - \mu \hat{N})}}{Z_\textnormal{lead}}.
\end{equation}
Here $\beta = (k_\textnormal{B} T)^{-1}$ and $Z_\textnormal{lead}$ is the grand-canonical partition function which ensures the normalization of the lead density matrix, $\textnormal{Tr}_\textnormal{lead} \hat{\rho}_\textnormal{lead} = 1$.
Typically, it is assumed that the initial state for the integration of the Liouville-von Neumann equation is separable, $\hat{\rho}(0) = \hat{\rho}_\textnormal{sys} (0) \otimes \hat{\rho}_\textnormal{lead}$. Consequently, the entanglement between the system and the lead will appear as a result of the hybridization between them and manifests in the reduced density matrix dynamics at $t>0$.
Using the Nakajima-Zwanzig projector operator technique \cite{Nakajima1958, Zwanzig1960, Petruccione2002}, an integro-differential equation for the reduced density matrix of the system, $\hat{\rho}_\textnormal{red} = \mathcal{P}\hat{\rho}$, where $\mathcal{P} := \textnormal{Tr}\{\bullet\} \otimes \hat{\rho}_\textnormal{lead}$ can be found
\begin{equation}\label{eq:reduced_general_eq}
  \dot{\hat{\rho}}_\textnormal{red}(t) = \mathcal{L}_\textnormal{sys}  \hat{\rho}_\textnormal{red}(t) + \int_0^t \textnormal{d}s \mathcal{K}(t-s)  \hat{\rho}_\textnormal{red}(s).
\end{equation}
The kernel superoperator, $\mathcal{K}(t)$, is given by
\begin{equation}
\label{eq:ker_sup}
 \mathcal{K}(t) = \mathcal{P}\mathcal{L}_\textnormal{tun} \bar{\mathcal{G}}(t) \mathcal{L}_\textnormal{tun} \mathcal{P},
\end{equation}
and it is a combination of the Liouvillean superoperators, the projector and the propagator
\begin{equation}\label{eq:green}
 \bar{\mathcal{G}}(t)  = \exp[\mathcal{L}_\textnormal{sys} + \mathcal{L}_\textnormal{lead} + (1 - \mathcal{P})\mathcal{L}_\textnormal{tun} (1 - \mathcal{P})](t),
\end{equation}
where each Liouvillian superoperator in Eqs. \eqref{eq:reduced_general_eq}-\eqref{eq:green} is given by Eq. \eqref{eq:liouvillian} after replacement of $\hat{H}$ by $\hat{H}_i$.

Eq. \eqref{eq:reduced_general_eq} is formally exact, it contains memory effects and describes the system dynamics at all perturbative orders in the tunneling Hamiltonian.
Due to its convolution form, it is natural to work in Laplace space. Employing the final value theorem, it is easy to show that the steady-state reduced density matrix satisfies the equation
\begin{equation}
 \left[\mathcal{L}_\textnormal{sys} + \tilde{\mathcal{K}}(0)\right] \hat{\rho}^\infty_\textnormal{red} = 0,
\end{equation}
where $\hat{\rho}^\infty \equiv \hat{\rho}(t\rightarrow \infty)$ and
\begin{equation}
 \tilde{\mathcal{K}}(0) = \lim\limits_{\lambda \to 0^+}  \int_{0}^{+\infty} \textnormal{d}t \,{\rm e}^{-\lambda t} \mathcal{K}(t) .
\end{equation}
To investigate the time evolution, the Markov approximation is applied to Eq. \eqref{eq:reduced_general_eq}, thus  leading to an equation of motion where we have discarded memory effects
\begin{equation}\label{eq:markov}
 \dot{\hat{\rho}}_\textnormal{red}(t) = \left[ \mathcal{L}_\textnormal{sys} + {\mathcal{K}}_{\textnormal{Markov}} \right] \hat{\rho}_\textnormal{red}(t),
\end{equation}
with ${\mathcal{K}}_{\textnormal{Markov}} = \tilde{\mathcal{K}}(0)$.

\subsection{Sequential tunneling}

From Eq.~\eqref{eq:ker_sup} we infer that the Markov approximation applies if the smallest time scale is the memory time. The latter is obtained from the correlator of the thermalized bath and reads $\tau_{\rm corr} = \hbar\beta$. Therefore the condition $\tau_{\rm corr} \ll \tau$, with $\tau$ the characteristic evolution time of the system, translates into having the tunnelling rate $\Gamma \sim 1/\tau$ smaller than the thermal energy, $\hbar \Gamma \ll k_\textnormal{B}T$ The latter is in turn smaller than the charging energy, $E_c$. As the tunneling rate is proportional to powers of the modulus square of the tunneling matrix elements $t_{i\mathbf{k}\sigma}$in Eq. \eqref{eq:tunneling},  under the condition of weak tunnelling coupling, the Markovian kernel can be expanded perturbatively in a controlled way.
Here, we only retain the first non-vanishing term in this perturbative expansion, $\tilde{{\mathcal{K}}} \simeq \tilde{{\mathcal{K}}}^{(2)} + \mathcal{O}(\tilde{\mathcal{K}}^4)$.  
From Eq.~\eqref{eq:ker_sup}, this first contribution to the kernel therefore reads \cite{Petruccione2002}
\begin{equation}\label{eq:kernel_second}
 \tilde{\mathcal{K}}^{(2)}(0) = \mathcal{P} \mathcal{L}_\textnormal{tun} \tilde{\mathcal{G}}_0 \mathcal{L}_\textnormal{tun} \mathcal{P},
\end{equation}
with
\begin{equation}
 \tilde{\mathcal{G}}_0 = \lim\limits_{\lambda \to 0^+}  \dfrac{1}{\lambda - \mathcal{L}_\textnormal{sys} - \mathcal{L}_\textnormal{lead}},
\end{equation}
being the Laplace transform of the free propagator in the absence of tunneling from or to the bath. Note that the formal limit $\lambda \rightarrow 0^+$ is to be taken only at the end of the calculation.
The kernel in Eq.~\eqref{eq:kernel_second} corresponds to the so-called coherent sequential tunneling regime \footnote{All the odd terms vanish due to particle-number conservation, \textit{i.e.} $\mathcal{P} \mathcal{L}_\textnormal{tun}^{n} \mathcal{P} = 0$ if $n$ is odd.}. For coherent sequential tunneling, the time between each tunneling event corresponds to the largest time scale in the system and there cannot be coherence between individual tunneling events.

After integrating out the bath degrees of freedom in Eq. \eqref{eq:kernel_second} the second order kernel reads
\begin{align}\label{eq:kernel_second_2}
\tilde{\mathcal{K}}^{(2)}(0) &= -\lim\limits_{\lambda \to 0^+}  \dfrac{i}{2\pi} \sum_{\alpha, \alpha'} \sum_{p, \sigma} \sum_{i,j} \int \textnormal{d} E \,\alpha \alpha' \Gamma^{p}_{ij}(E, \sigma)  \notag \\
&\times \mathcal{D}^{\bar{p}}_{i\sigma \alpha'} \dfrac{f^{(p\alpha)}(E)}{p E - i \hbar \mathcal{L}_\textnormal{sys} + i\lambda} \mathcal{D}^p_{j\sigma \alpha},
\end{align}
where we have used
\begin{equation}
\textnormal{Tr}\left(c^{p,\alpha}_{\mathbf{k}\sigma} c^{p',\alpha'}_{\mathbf{k}'\sigma'}\hat{\rho}_\textnormal{lead} \right) = \delta_{\mathbf{k}\mathbf{k}'} \delta_{\sigma \sigma'} \delta_{p\bar{p}'} f^{(p \alpha')}(E).
\end{equation}
Here, the Liouville indices $\alpha, \alpha' = \pm 1$, $\mathcal{D}_\alpha$  are superoperators defined by their action from the left/right (resp. $+/-$) on the density matrix
\begin{align}
 \mathcal{D}^{p}_{i\sigma+} \hat{\rho} &= \hat{d}_{i\sigma}^p \hat{\rho}, \\
 \mathcal{D}^{p}_{i\sigma-} \hat{\rho} &= \hat{\rho} \hat{d}_{i\sigma}^p,
\end{align}
and $f^+(E) = \{\exp[\beta(E-\mu)] + 1\}^{-1}$, $f^-(E) = 1 - f^{+}(E)$ are Fermi distribution functions.

The geometry and the hybridization of the single-particle defect states with the fermionic reservoir are contained in the single particle (tunneling) rate matrix (also known as tunneling self energy) \cite{Donarini2019, Donarini2021}
\begin{equation}\label{eq:gamma}
 \Gamma_{ij}^p(E, \sigma) = \dfrac{2 \pi}{\hbar} \sum_\mathbf{k} t^{\bar{p}}_{i \mathbf{k} \sigma} t^{p}_{j \mathbf{k} \sigma} \delta(\epsilon_{\mathbf{k}\sigma}-E),
\end{equation}
where $\Gamma^+_{ij}(E, \sigma) = \Gamma^{-}_{ji}(E,\sigma)$.
This matrix extends the concept of tunneling rate and  its form determines the strength of the interference effects between quasi-degenerated many-body states \cite{Donarini2009,Wacker2011}. The matrix $\bm{\Gamma}(E)$ is defined in the relevant sector of the single-particle Hilbert space, spanned in this study by the two spinfull empty levels of the XS\textsubscript{2} vacancy.
The calculation of the rate matrix, combining \textit{ab initio} results and the symmetry analysis from from Sec. III in the SI, is now performed in Sec. \ref{sec:results}.

\section{Relaxation rates and dynamics}\label{sec:results}

\begin{figure}
    \includegraphics[width=0.825\linewidth]{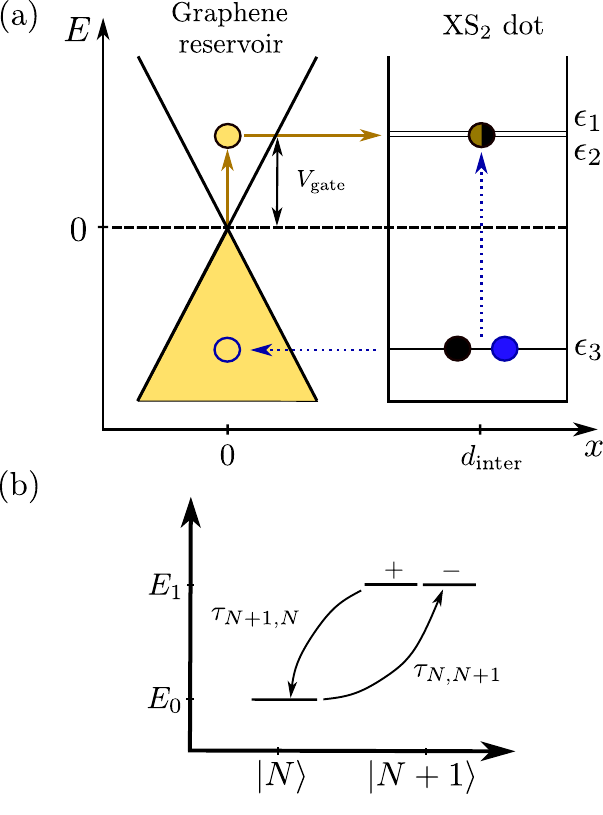}      
    \caption{(a) Sketch of the low energy landscape of the XS\textsubscript{2} dot (here X = W) and the graphene reservoir, providing a structured DoS. 
    We also show the potential single-particle electronic transitions occurring within or to the vacancy.
    The graphene reservoir is separated by a distance $d_\textnormal{inter}$ from the dot-containing ``active'' layer. The position of the vacancy energy levels $\epsilon_i$ are determined with respect to the graphene charge neutrality point ($\mu = E_\textnormal{F}$, set as the zero of energies). We represent two possible pathways for a $N \rightarrow N+1$  transition: (i) a photoexcited electron is promoted to one of the two empty quasi-degenerate levels leaving behind a hole in the originally doubly occupied state of the dot, which can be refilled with the hole tunneling towards graphene (blue dotted arrows) (ii) by means of an electric field created by gate voltage or doping the chemical potential of graphene can be changed and an electron tunnel from the graphene layer into one of the empty vacancy levels (yellow solid arrows). 
    (b) Representation of the many-body Hilbert space and the transition times, $\tau_{\alpha, \beta}$, between the many-body initial state $\alpha$ and final state $\beta$ (characterized by the total particle number $N,N+1$) in the absence or presence of very weak spin-orbit coupling. The many-body energies are considered in the grand canonical ensemble, $E_0 = \varepsilon_0 -N \mu$ and $E_1 = \varepsilon_1 -(N+1)\mu$, with $\mu \equiv E_\textnormal{F}$ at zero gate voltage
    and $\varepsilon_{0,1}$ are the total energies.
    }\label{f8} 
\end{figure}

We now discuss the relaxation rates and kinetic equations for our minimal model. As $E_c$ is the larger energy scale, we consider only the addition of a single electron to the vacancy. Therefore, the interaction term in Eq. \eqref{eq:hamiltonian_sys} does not contribute explicitly to the transition rates.
We denote the ground state of the system with an empty vacancy by $|N \rangle$; this many-body state corresponds to a state in which two levels of the impurity are empty and the one far below the Fermi level is fully occupied, as in the equilibrium situation described by DFT. 
Correspondingly, the charged system in which an additional electron has tunneled into the vacancy is represented by the many-body state $|N+1 \rangle = \hat{d}^\dagger_{i\sigma} | N\rangle$ with one additional electron created at the impurity level described by the quantum numbers $(i,\sigma)$. 

We envisage two possible pathways leading to a singly charged impurity, as shown schematically in Fig. \ref{f8} (a). In the first case, a photoexcited electron from the occupied dot levels is promoted to one of the empty levels of the vacancy. Due to hybridization of the defect states and the valence band at the XS\textsubscript{2} layer, a hole subsequently tunnels towards the graphene layer. Since this process involves photoexcitation of the electron (photoemission for the reverse process) it naturally involves excitons. We shall not consider it here, for the sake of simplicity.
The empty dot levels can by populated in a second way by increasing the chemical potential of the graphene layer by means of an external back gate or by doping. In this case, electron tunneling occurs directly from the reservoir to the empty single-particle level at the vacancy.
The reverse process would occur by tunneling from the previously occupied vacancy states into the graphene layer in a similar way.

\subsection{Weak spin-orbit interaction}\label{sec:weak}

\subparagraph{Single particle rate matrix.} 
If the  spin-orbit interaction is weaker than the coupling to the reservoir, as it is for MoS\textsubscript{2}, we can, at first, neglect $\hat{H}_{\rm SOC}$ and concentrate on the single-particle Hamiltonian including only the degenerate transport levels, the tunnelling term and the reservoir, \textit{e.g.}, the two empty defect ``bands'' tunnel-coupled to the graphene Dirac cone. We will apply the same analysis also to the WS\textsubscript{2}--Gr heterostructure, this time to better highlight  the effects of spin-orbit interaction, presented later in Sec. \ref{sec:soc}. 
In both cases, the relevant Fock space has $\textnormal{dim}\,\mathcal{F} = 5$, with the $N$-sector (vacuum) being one-dimensional (as it corresponds to a closed-shell ground state) and the $N+1$-sector four-dimensional (due to spin and orbital degrees of freedom). Since the $N+1$-sector is a simple single-particle sector, its dimension coincides with the one of the (single-particle) Hilbert. 
Consequently, $\bm{\Gamma}$ is a four-dimensional matrix which, in the absence of spin-orbit interaction, simply factorizes into an orbital and a trivial spin component: $\boldsymbol{\Gamma} = \boldsymbol{\Gamma}_{\rm orb} \otimes \bm{1}_2$
Moreover,  we show in Sec. III in the SM that the  orbital component of the single particle rate matrix is also diagonal with the form
\begin{equation}\label{eq:gamma_uu}
 \Gamma^{\rm orb}_{u u'} (E)= \dfrac{\pi^2}{2\hbar}  \dfrac{A_\textnormal{sc}}{A_\textnormal{c}} |t_u|^2 \rho_\textnormal{Gr}(E)\delta_{u u'},
\end{equation}
where $u$ and $u'$ are the quantum number characterizing the symmetric/antisymmetric defect states, $A_{sc}$ is the area of the supercell and the DoS of graphene is given by  $\rho_\textnormal{Gr}(E) = 2 A_c |E|/(\pi \hbar^2 v_\textnormal{F}^2)$ \cite{Geim2008}. Here, we considered the unit cell area of graphene given by $A_c =3 \sqrt{3} a^2/2$ with $a = 1.42$ \AA\, the graphene Fermi velocity $v_\textnormal{F} \sim 10^6 \,\textnormal{m}/\textnormal{s}$ and the ratio $A_\textnormal{sc}/A_\textnormal{c} = 25$. Thus, $\boldsymbol{\Gamma}^{\rm orb}$ can also be cast into the form
\begin{equation}
 \bm{\Gamma}^{\rm orb} = \dfrac{\Gamma_{uu} + \Gamma_{\bar{u}\bar{u}}}{2} \bm{1}_2 + \dfrac{\Gamma_{uu} - \Gamma_{\bar{u}\bar{u}}}{2} \boldsymbol{\tau}_z,
\end{equation}
Note that this object has dimensions of $\textnormal{[time]}^{-1}$.

The modulus of the hybridization matrix elements can be directly extracted from the DFT band structure  by looking at the anti-crossing of the energy levels close to $\bar{\textnormal{K}}$,  considering this problem as a local two-level problem in $\mathbf{k}$-space with Hamiltonian
\begin{equation}\label{eq:local_ham}
 \hat{H}_\textnormal{loc} = 
 \begin{pmatrix}
  E_1 & t \\
  t^\ast & E_2
 \end{pmatrix}.
\end{equation}
The straightforward diagonalization of Eq. \eqref{eq:local_ham} gives the eigenenergies
\begin{equation}\label{e2}
 E_{\pm} = \dfrac{E_1 + E_2}{2} \pm \dfrac{1}{2} \sqrt{(E_1 - E_2)^2 + 4|t|^2}.
\end{equation}
In the literature \cite{Hongyi2017, Malic2020}, this expression is sometimes known as ``avoided crossing formula''. We then estimate $|t|$ by choosing the zero of energy at the point where the unperturbed energies are degenerated, $|t| = (E_{+} - E_{-})/2$. 
From the DFT band structures [Figs \ref{f2db} (a) and (c)]  we find $|t^{\rm Mo}_u| = 12 $ meV, $|t^{\rm Mo}_{\bar{u}}| = 8$ meV, together with $E_u = 57$ meV for MoS\textsubscript{2}--Gr and $|t_u^{\textnormal{W}}| = 15$ meV, $|t_{\bar{u}}^{\rm W}| = 5$ meV with $E_u = 298$ meV for WS\textsubscript{2}--Gr.
%
%

\subparagraph{Transition rates.}
Using Eq. \eqref{eq:kernel_second_2}, we can obtain the transition rates for the charging process between an initial uncharged and final charged vacancy, $N \rightarrow N +1$, as well as the reverse - discharging - process $N +1 \rightarrow N $, in the lowest non vanishing perturbative order and as shown in Fig. \ref{f8} (b). 
As $\Gamma_{ij}(E, \sigma)$, as well as the Fermi function $f^{(p\alpha)}(E)$ are smooth functions, we can evaluate the integral by using the Sokhotski-Plemejl theorem over the real line \cite{Niklas2018}
\begin{equation}\label{eq:SP_theorem}
  \lim\limits_{\lambda \to 0^+}   \int_{-\infty}^{+\infty} \textnormal{d}\omega \dfrac{h(\omega)}{\omega + i \lambda} =  \textnormal{p.v.}  \int_{-\infty}^{+\infty} \textnormal{d}\omega \dfrac{h(\omega)}{\omega} -i \pi h(0), 
\end{equation}
where p.v. stands for principal value and $h(\omega)$ is a complex-valued function  \footnote{The finite band width of the graphene DoS ensures the convergence of the principal value integral.}   
From the real part of $\tilde{\mathcal{K}}^{(2)}(0)$ acting on the reduced density matrix $\hat{\rho}_\textnormal{red}$ we extract the transition rate $W_{\alpha,\beta}$ between an initial $\alpha$ and final $\beta$ many-body state. The projection on the many-body Hilbert space of the vacancy yields
  \begin{align}\label{e7}    
  W_{N,N+1}  &= \sum_u W_{0\rightarrow u}, \notag \\
 &= \sum_{u\sigma} \int dE  \langle N | \hat{d}_{u\sigma} |N+1  \rangle \langle N+1| \hat{d}_{u\sigma}^\dagger | N \rangle  \notag \\ &\times \Gamma_{u}(E)f^+(E) \delta(E -\Delta E_{0, u}),
  \end{align}
where $\Gamma_u:=\Gamma_{u{u}}$ and by definition $\Delta E_{0,u} = E_u - E_0 > 0$. This expression simplifies to
\begin{equation}\label{eq:rate_01}
 W_{N, N+1} = 2 \sum_u \Gamma_{u}(\Delta E_{0,u}) f^+(\Delta E_{0, u}),
\end{equation}
 where the factor of two stems from the spin degeneracy. 
From this rate, we define the transition time for the population of the vacancy by an additional electron as $\tau_{N , N+1} := 1/W_{N , N+1}$.
The discharging process in which the additional electron initially present at the vacancy leaves the dot by tunneling back towards the graphene layer is given by
 \begin{equation}\label{eq:rate_10}
  W_{N+1, N} = 2 \sum_u\Gamma_{u}(\Delta E_{0, u}) f^-(\Delta E_{0, u}).
 \end{equation}
The corresponding transition time can be defined analogously as for the charging process, $\tau_{N+1, N} := 1/W_{N+1, N}$.

Evaluation of Eqs.~\eqref{eq:rate_01}-\eqref{eq:rate_10} at room temperature ($T=300$ K) and chemical potential $\mu = E_\textnormal{F}$ ($\mu \equiv 0$ at the charge neutrality point) yields $\tau^{\rm Mo}_{N, N+1}  = 29.2$ ps and $\tau^{\rm Mo}_{N+1, N}  = 3.1$ ps for the MoS\textsubscript{2}--Gr heterobilayer and $\tau^{\rm W}_{N, N+1}  = 47.1$ ns and $\tau^{\rm W}_{N+1, N}  = 0.5$ ps for the WS\textsubscript{2}--Gr heterobilayer. The charging and discharging tunnelling times are crucially influenced by the number of states in the graphene reservoir available to tunnel at the resonant energy. Ultimately, the Dirac dispersion of graphene (linear DoS) and the Pauli exclusion principle (Fermi function) set their values.

Thus, the substantially longer charging time for the case of WS\textsubscript{2} compared to MoS\textsubscript{2} can be understood from the higher energy difference between the empty vacancy states and the chemical potential of graphene in the case of WS\textsubscript{2} as compared to the ones of MoS\textsubscript{2} [see Figs. \ref{f2db} (a) and (c)]. Even at room temperature and despite the larger DoS in graphene, only few thermally excited electrons can populate the impurity.
On the contrary, the discharging time is shorter for WS\textsubscript{2} compared to MoS\textsubscript{2}. This is a consequence of the larger amount of available states to tunnel to in graphene for the WS\textsubscript{2}--Gr interface, which manifests itself in the density of states, as the Pauli exclusion priciple cannot hinder the population of empty states.
We note that graphene being a semimetal leads to a counter intuitive behavior of the transition times compared to a metallic reservoir because the closer the state is to the Dirac point, the less available states exists to tunnel from and to.

The calculated relaxation time at zero chemical potential roughly agree to the one experimentally found in tr-ARPES experiments, in which the electron is photoexcited to a state in the conduction band of the TMDC layer \cite{Gierz2020, Gierz2021}. In the experiment, electronic lifetimes of the order of $\sim 1$ ps are measured from the pump-probe signal for the WS\textsubscript{2}--Gr heterostructure. 
Though, a direct comparison between our calculations and these experimental lifetimes is \textit{strictu senso} not easy, because in this experiment, first excitons are resonantly excited as initial state for the dynamics of hot carriers. Although thermalization and exciton dissociation are expected to occur at shorter timescales $ \lesssim 100 $ fs, compatible with a binding energy of $\sim 200$ meV found in recent calculations for the ``A-like'' exciton \cite{Kleiner2022}, the exciton dissociation and recombination to the defect states can strongly influence the dynamics.
Second, the measured dynamics would involve also the pristine conduction (or valence, in the case of hole transfer) bands, a scenario different from what we are considering here, where those bands are assumed to be far in energy from the relevant defect transport channels and therefore do not participate in the dynamics. Finally, our theory considers the extremely dilute limit in which the vacancies are relatively isolated, this scenario may change in the experiment and the charge transfer rates differ if the vacancy concentration is higher. 
In any circumstance, the defect associated transport channel acts as a mechanism that slows down the charge transfer.
%
%
%
%
\begin{figure}
    \includegraphics[width=0.95\linewidth]{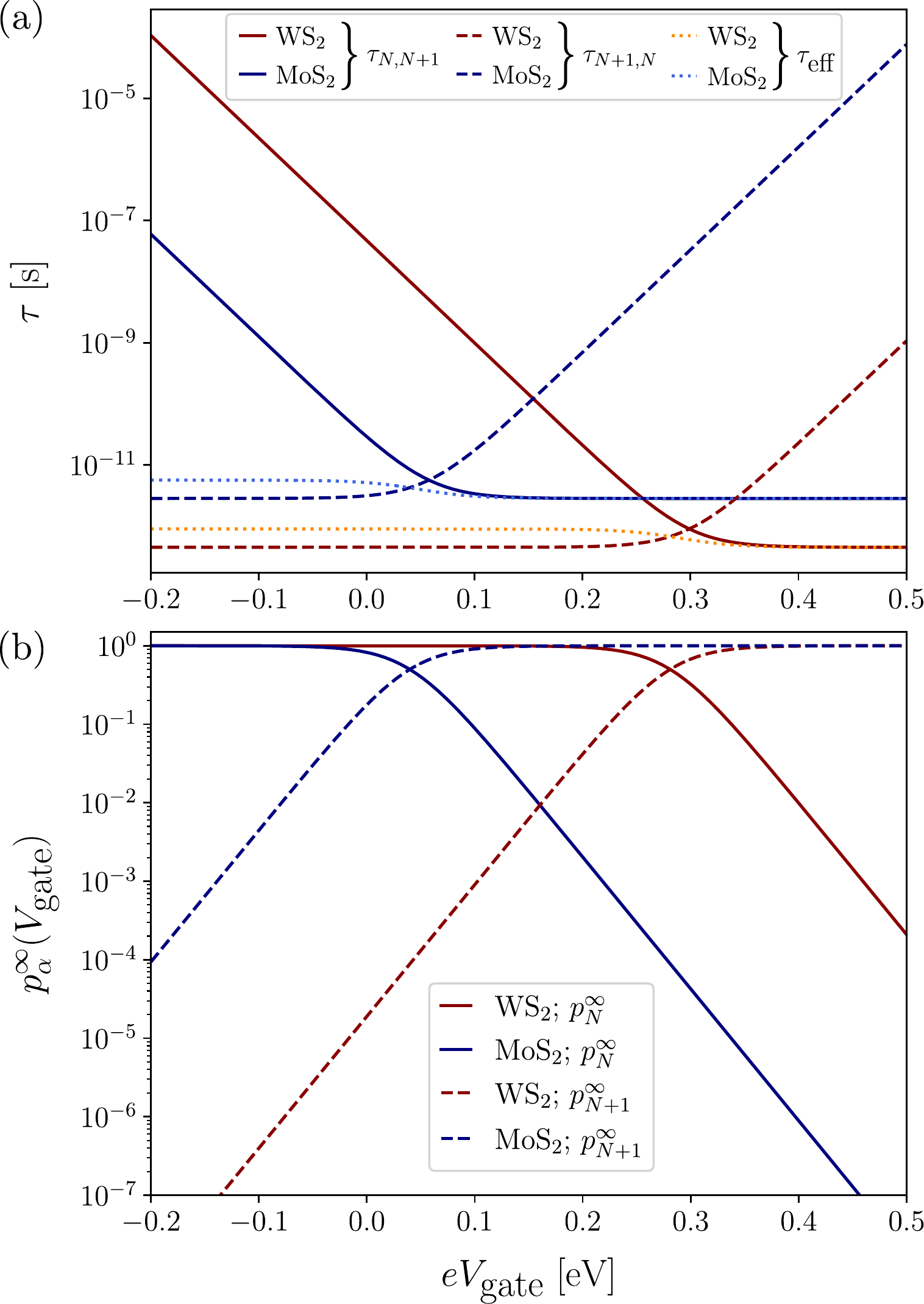}
    \caption{(a) Electronic transition times, $\tau$, for MoS\textsubscript{2} and WS\textsubscript{2} vacancy states as a function of the gate voltage, $V_\textnormal{gate}$ measured in eV. The solid and dashed lines correspond to the transitions $\alpha \rightarrow \beta$ with $\alpha, \beta \in \{N,N+1\}, \alpha \neq \beta$ as defined in Eqs. \eqref{eq:rate_01} and \eqref{eq:rate_10}. The dotted line corresponds to the effective transition time associated to the semiclassical dynamics defined from Eq. \eqref{eq:rate_eff} and controlling the dynamics from a local to global equilibrium.
    For simplicity, we set the chemical potential at zero gate voltage to be zero. 
    (b) Corresponding stationary semiclassical probability of occupation of the many-body levels, $\mathbf{p}^\infty$, as a function of the gate voltage. As anticipated, the stationary populations follow the Fermi distribution function.
    }\label{f10} 
\end{figure}
%
%

Bath temperature, electric fields or changes in the position of the single-particle defect energies due to screening can have a dramatic effect in the transition rates.
We analyze here the behavior of the transition rates with the electric field created by a voltage drop in the direction perpendicular to the heterostructure (the analysis of the effect of temperature and defect level energy shifts due to screening is given in Sec. IV of the SI).
In Fig. \ref{f10} (a) we display the transition rates as a function of the gate voltage, $eV_\textnormal{gate}$, which may be tuned by an external gate located under the graphene layer and which creates a voltage drop between graphene and TMDC layers. 
The window of voltage drop between the two layers considered here corresponds to a range of electric fields between the layers of the order of $E \lesssim 3$ V/nm. This is consistent with electric fields considered in previous theoretical works \cite{Fabian2019} and it is substantially smaller than the experimental dielectric breakdown points of these materials \cite{Eroms2022}.
We first observe that for each material whether a charging ($N \rightarrow N+1$) occurs faster compared to a discharging ($N+1 \rightarrow N$) process depends on the value of the chemical potential ($\simeq$ gate voltage) with respect to the position of the defect levels. This can be easily understood from the functional form of $f^+(x)$ and $f^-(x)$ in Eqs. \eqref{eq:rate_01}-\eqref{eq:rate_10}: for $\mu < \Delta E_{0, u}$, charging is slower because the tail of the Fermi function (that has a typical width of a few $k_BT$) is small and there are no occupied states that can tunnel towards the vacancy, while for $\mu > \Delta E_{0, u}$ the reverse situation occurs and discharging becomes slower because there are very few (depending on temperature) empty states in graphene capable to receive electrons tunnelling from the vacancy.
The charging and the discharging only have the same transition times and rates if the chemical potential aligns precisely with the defect energies, $\mu = \Delta E_{0,u}$ (thus $\tau_{N, N+1} = \tau_{N+1, N}$). This is a trivial consequence of $f^+( \Delta E_{0,u}) = f^-( \Delta E_{0,u})$, together with the fact that $\bm{\Gamma}$ is a diagonal matrix (therefore $\bm{\Gamma}^+ = \bm{\Gamma}^-$).

We now discuss the differences between the two TMDCs in the XS\textsubscript{2}--Gr interface. We analyze the case $\mu \gg \Delta E_{0,u} $, the scenario for $\mu \ll \Delta E_{0,u}$ can be deduced with the mapping $\tau_{N, N+1} \leftrightarrow \tau_{N+1, N}$ upon reversing the inequality. If the chemical potential is substantially larger than the defect energy, $\mu \gg \Delta E_{0,u}$, we observe that $\tau_{N, N+1}$ saturates and the charging is faster for X = W compared to X = Mo. This is a consequence of the WS\textsubscript{2}--Gr heterostructure having larger single-particle tunneling rate compared to MoS\textsubscript{2}--Gr, mainly due to the larger number of states in graphene to tunnel from in the former compared to the latter [$\rho_\textnormal{Gr}(\Delta E^\textnormal{W}_{0, u}) > \rho_\textnormal{Gr}(\Delta E^{\textnormal{Mo}}_{0, u})$].
The differences in the charging and descharging rates between the two heterobilayers can be tuned with the gate voltage. For example, for $\mu >0$, there is a voltage threshold for which the relative ordering between the charging times of the two materials changes [in other words, the continuous lines in Fig. \ref{f10} (a) cross]. This point, $\mu_\textnormal{cr}$, marks the change from $\tau_{N, N+1}(\textnormal{W})>\tau_{N, N +1}(\textnormal{Mo})$ if $\mu < \mu_\textnormal{cr}$ to $\tau_{N, N+1}(\textnormal{W})<\tau_{N, N +1}(\textnormal{Mo})$ if $\mu > \mu_\textnormal{cr}$.
Though their difference can be modulated, in the discharging, WS\textsubscript{2} shows a smaller transition time $\tau_{N+ 1, N}$ (faster discharging) compared to MoS\textsubscript{2}, independently of the chemical potential. This result is attributed to the higher energy of the defect state and to its lower tunneling barrier to the graphene layer for X = W as compared to X = Mo.

\subparagraph{Semiclassical dynamics.} 
The kinetic equation Eq.~\eqref{eq:markov} describes a time-local dynamics for the vacancy in presence of the graphene bath still involving quantum coherences between quasi-degenerate vacancy states. The latter are captured by the coherences ({\it i.e.} the off-diagonal elements of the density matrix $\hat{\rho}_{\rm red}$, written in the energy eigenbasis). Starting from Eq.~\eqref{eq:markov} it is easy to derive the semiclassical Pauli master equation only involving the probability of occupation of a given many-body state \cite{Bruus2004}, written in terms of the charging and discharging rates, already discussed in the previous section.
Thus, the probability, $p_\alpha := p_\alpha(t)$, of finding the heterostructure in the state $\alpha$ with given particle number at any time $t$, is obtained from the solution of
\begin{equation}\label{eq:populations}
\dot{p}_{\alpha} = \sum_\beta \left[p_\beta W_{\beta, \alpha} -p_\alpha W_{\alpha, \beta} \right] ,
\end{equation}
together with the conservation of probability, $\sum_\alpha p_\alpha = 1$ [which results from $\textnormal{Tr}(\hat{\rho}) = 1$] and a given initial condition. We further assume a fast local thermalization between the tunnelling events or local  thermal equilibrium, so that the population of the degenerate energy levels has to be the same, $p_u = p_{\bar{u}}$. As a consequence, we reformulate Eq. \eqref{eq:populations} as a two-level problem characterized by the populations $p_N$ and $p_{N+1} = p_{u} + p_{\bar{u}}$ in order to investigate the dynamics towards a final state of global equilibrium of the vacancy in contact with the thermal bath.
The analytical solution of the associated system of equations gives the vector of probabilities
\begin{equation}\label{eq:rates_t}
 \mathbf{p}(t) = \mathbf{p}^\infty + {p}_0
 \begin{pmatrix}
  1 \\
  -1
 \end{pmatrix}
 {\rm e}^{- W_{\rm eff} t}.
\end{equation}
We therefore find that the vacancy relaxes from an initial state following an expected exponential decay law typical for a tunneling process with and effective rate 
\begin{equation}\label{eq:rate_eff}
    W_\textnormal{eff} := W_{N, N+1} + \tilde{W}_{N+1, N},
\end{equation}
where $\tilde{W}_{N+1,N} := W_{N+1,N}/2$. This effective rate defines an effective relaxation time $\tau_\textnormal{eff} := 1/W_{\textnormal{eff}}$ for the population dynamics of the isolated vacancy. Substitution using Eqs. \eqref{eq:gamma_uu}, \eqref{eq:rate_01} and \eqref{eq:rate_10} gives $\tau^\textnormal{W}_\textnormal{eff} = 0.9$ ps and  $\tau^\textnormal{Mo}_\textnormal{eff} = 5.2$ ps, as the time scale at which the system reaches equilibrium. This value is of the same order of magnitude of the electronic relaxation times obtained from tr-ARPES measurements. We show $\tau_\textnormal{eff}$ in Fig. \ref{f10} (a) for X = W, Mo as a function of $V_\textnormal{gate}$. In comparison to $\tau_{N, N+1}$ and  $\tau_{N+1, N}$, we find that this relaxation rate is a smoother function of the voltage, roughly one order of magnitude larger in MoS\textsubscript{2} compared to WS\textsubscript{2}.

 \begin{figure}
    \includegraphics[width=1.0\linewidth]{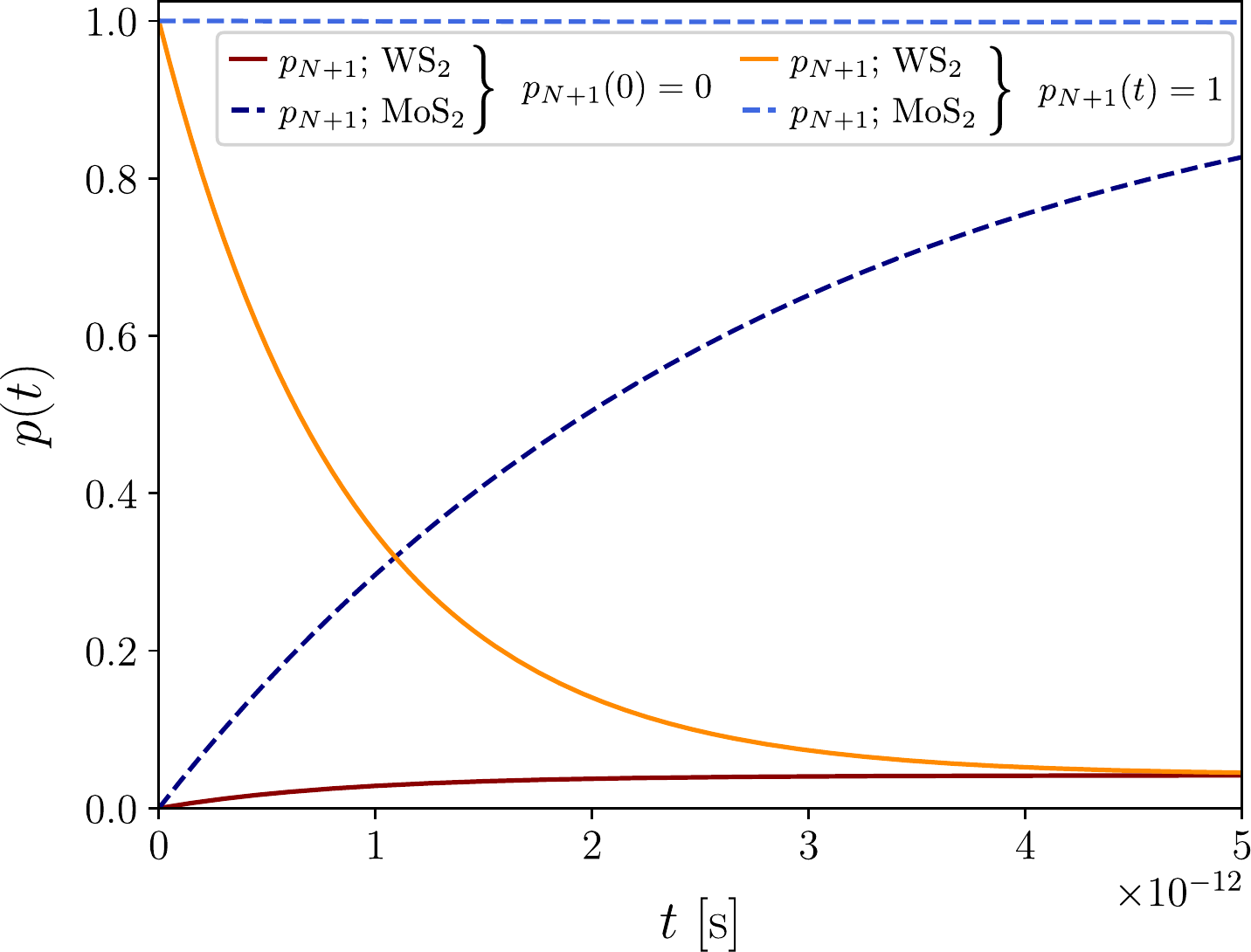}
    \caption{Population, $p(t)$, of the many-body levels as a function of time, $t$ (measured in s). The dynamics is computed for the representative gate voltage, $V_\textnormal{gate} = 0.2$ eV. We only show the population for the state $|N+1 \rangle$, the population of the state $|N \rangle$ being obtained as $p_N(t) = 1 - p_{N+1}(t)$.
    }\label{f16} 
\end{figure}

The vector $\mathbf{p}^\infty$ corresponds to the stationary populations of the states $|N \rangle$ and $|N+1 \rangle$
\begin{equation}
 \mathbf{p}^\infty = 
 \begin{pmatrix}
  p_N^{\infty} \\
  p_{N+1}^{\infty} 
 \end{pmatrix}
 = \dfrac{1}{W_\textnormal{eff}}
 \begin{pmatrix}
  \tilde{W}_{N+1, N} \\
   W_{N, N+1}
 \end{pmatrix}.
\end{equation}
This object only depends on the ratio of many-body transitions rates. %
Note that, as expected for second order processes, the ratio of the equilibrium populations is proportional to the Boltzmann factor \cite{Petruccione2002}. Indeed, it is easy to check that this ratio is $p^{\infty}_{N+1}/p^\infty_N = g_{N+1} \exp(-\beta \Delta E_{0,u})$ where $ g_{N+1} = 2$ is the (Boltzmann) degeneracy of the state $|N+1 \rangle$ (by construction, $g_N = 1$). The prefactor $p_0$ in Eq. \eqref{eq:rates_t} depends on the initial condition for the time evolution. If the system initially had $N$ electrons, then, $p_0 = W_{N, N+1}/W_\textnormal{eff}$, while if the system initially was in the many-body state with particle number equal to $N+1$ it reads $p_0 = - W_{N+1, N}/W_\textnormal{eff}$.

The stationary many-body populations for each of the two heterostructures  as a function of the gate voltage, $V_\textnormal{gate}$, and computed with realistic \textit{ab initio} parametrization are shown in Fig. \ref{f10} (b). 
As expected, the stationary population follows a Fermi distribution function since at $t \rightarrow +\infty$ the system has relaxed to the equilibrium distribution provided by the thermal bath.
The values of the populations are controlled by the transition rates only and shifted by the difference between the defect energies for each material. In the large positive $eV_\textnormal{gate}$ limit, we have $p_N^\infty \rightarrow 0$ and $p_{N+1}^\infty \rightarrow 1$ as the transition time associated to the charging of the vacancy becomes smaller and saturates, as seen in Fig. \ref{f10} (a). Due to the symmetry relation $\tau_{N, N+1} \rightarrow \tau_{N+1, N}$ with respect to the defect energy, the situation is reversed for large negative $V_\textnormal{gate}$.  

In Fig. \ref{f16} we show an example of the population dynamics for a given voltage gate, $V_\textnormal{gate} = 0.2$ V, following Eq. \eqref{eq:rates_t}. We only display the population of the state $|N+1 \rangle$, for two initial conditions, $p_{N+1}(0) = 0$ and $p_{N+1}(0) = 1$. The population shows the expected exponential decay law, controlled by $W_\textnormal{eff}$, as well as the asymptotic limit given in Fig. \ref{f10} (b) which corresponds to the situation of global thermodynamic equilibrium. Note that properly chosen combinations of material and $V_\textnormal{gate}$ can strongly reduce the charge relaxation, also depending on the initial conditions, effectively inhibiting charge quenching at the impurity, as it occurs here for the initially charged vacancy at the MoS\textsubscript{2}--Gr interface (see light blue dashed curve).

\subparagraph{Lamb shift and quantum internal dynamics.} 

The imaginary part of $\tilde{\mathcal{K}}^{(2)}$ in Eq. \eqref{eq:kernel_second_2} acting on the reduced density matrix $\hat{\rho}_\textnormal{red}$, gives the so-called Lamb shift correction \cite{Donarini2009, Braun2004}. This term induces an effective internal dynamics due to a renormalization of the system Hamiltonian generated by virtual fluctuations of the particle number (which is conserved in average  \cite{Niklas2018}). The Lamb shift correction is only relevant for quasi-degenerate states and, in presence of a single bath, it can not influence the stationary state, thus influencing, in general, only the short time dynamics of the system.
For its derivation, one proceeds as in Ref. \cite{Donarini2021} to compute the principal value of the integral from Eq. \eqref{eq:SP_theorem}, where a particular attention should be given to the energy dependence of the density of states in graphene.
The second order kernel \eqref{eq:kernel_second_2} is therefore split into the tunneling and Lamb-shift terms $\tilde{\mathcal{K}}^{(2)} = \tilde{\mathcal{K}}^{(2)}_{\textnormal{T}} + \tilde{\mathcal{K}}^{(2)}_{\textnormal{LS}}$
and the equation of motion \eqref{eq:markov} can be rewritten in the form
\begin{equation}
    \dot{\hat{\rho}}_\textnormal{red} = -\dfrac{i}{\hbar}\left[ \hat{H}_\textnormal{sys} + \hat{H}_\textnormal{LS}, \hat{\rho}_\textnormal{red} \right] + \tilde{\mathcal{K}}^{(2)}_\textnormal{T} \hat{\rho}_\textnormal{red},
\end{equation}
where the commutator contains the Lamb-shift Hamiltonian, which, in the one particle sector, reads
\begin{equation}
\label{eq:H_LS}
 \hat{H}_{\textnormal{LS}}  = \dfrac{\hbar}{2\pi}  \sum_{u \sigma}  \left[\hat{d}_{u\sigma} \Upsilon_u(\hat{H}_{\rm sys})\hat{d}_{u\sigma}^\dagger + \hat{d}_{u\sigma}^\dagger \Upsilon_u(\hat{H}_{\rm sys})\hat{d}_{u\sigma}  \right].
\end{equation}
In the previous equation we have introduced the function
\begin{equation}
\Upsilon_u(\hat{H}_{\rm sys}) = {\rm p.v.}\int dE \frac{\Gamma_{u}(E)f^+(E)}{E-\hat{H}_{\rm sys} + {E}_{\rm av}},
\end{equation}
where $E_{\rm av} = (E_u + E_{\bar{u}})/{2}$.

As anticipated above, the dynamics induced by the Lamb shift Hamiltonian only influences, in the presence of a single bath, the short time scales. It produces pseudo-exchange fields leading to a ``precession-like'' dynamics in the quasi-degenerate subspace \cite{Donarini2021, Donarini2021b}. The latter is particularly relevant in ultrafast pump-probe experiments. Here we concentrate, instead, on a semiclassical dynamics, as we always assume local thermal equilibrium, thus leaving the full analysis of the coherent dynamics of the impurity quasi-degenerate states for future work.

\subsection{Strong spin-orbit interaction}\label{sec:soc}
\subparagraph{Single-particle rate matrix.}

In order to include spin-orbit interaction in our model, we could start from  the superposition of the atomic contributions \cite{Donarini2021b}
\begin{equation}\label{eq:LS}
 \hat{H}_\textnormal{SOC} = \sum_\alpha \xi_\alpha \hat{\mathbf{L}}_{\alpha} \cdot \hat{\mathbf{S}}_\alpha,
\end{equation}
where the summation runs over all atoms and orbitals, with $\hat{\mathbf{L}}_\alpha$ being the angular momentum operator and $\hat{\mathbf{S}}_\alpha$ the spin operator. 
Instead, we take advantage of the fact that the atomic spin-orbit interaction is already incorporated by construction into the Kohn-Sham Hamiltonian and project its spin-orbit contribution onto the vacancy Kohn-Sham empty orbitals that form the relevant Hilbert space discussed in Sec. \ref{sec:weak}.
The local C\textsubscript{3v} symmetry of the vacancy implies that the degenerate orbitals are associated to the quasi-angular momenta $L_z = \pm 1$.  From the decomposition
\begin{equation}
    \mathbf{L} \cdot \mathbf{S} = \dfrac{1}{2} \left[L_{+} S_{-} + L_{-} S_{+} \right] + L_z S_z,
\end{equation}
with $L_{\pm} = L_x \pm i L_y$ (similarly $S_{\pm}= S_x \pm i S_y$), we thus obtain that only the diagonal term containing $L_z$ can contribute to the projection. The resulting effective Hamiltonian is then characterized by a single spin-orbit parameter, $\xi_\textnormal{eff}$, and reads
\begin{equation}\label{eq:soc_ham}
 \hat{H}_\textnormal{SOC} = \frac{\xi_\textnormal{eff}}{2} \sum_{\ell,\sigma = \pm} \ell\sigma \, \hat{d}^\dagger_{\ell \sigma} \hat{d}_{\ell \sigma}.
\end{equation}
Here $\ell$ indicates the quasi angular momentum of the vacancy state, while the second index corresponds to the spin $\sigma = \pm$.

The value for the effective spin-orbit parameter can be estimated from the band structure looking at the spin-orbit splitting at the $\bar{\Gamma}$ point. From Fig. \ref{f6} we find $\xi^{\textnormal{Mo}}_\textnormal{eff} = 0.05 $ eV and $\xi^{\textnormal{W}}_\textnormal{eff} = 0.168 $ eV. Evidently, due to the heavier nature of the W atoms compared to Mo, the effective spin-orbit parameter in WS\textsubscript{2}--Gr is $\sim 3.5$ times larger compared to the MoS\textsubscript{2}--Gr heterostructure.
At the $\bar{\Gamma}$ point, where the impact of the impurity band dispersion is strongly reduced and the system displays the behavior of the isolated defect, we find that the distribution of the vacancy levels with spin-orbit interaction reads $E_{\ell,+} = E_{\bar{\ell}, -}$ and  $E_{\ell,-} = E_{\bar{\ell}, +}$. The Kramers' degeneracy is preserved and  the spin-orbit coupling has opposite effect for each of the $\ell, \bar{\ell}$ projections: $E_{\ell\sigma} = E_\textnormal{av} + \ell\sigma \xi_\textnormal{eff}/2$ with $E_\textnormal{av}$ the reference mid-gap energy.
For simplicity, we will use the short-hand notation $g := \{ \ell-,\bar{\ell}+ \}$ and $e := \{ \ell+,\bar{\ell}- \}$, with $E_e > E_g$ in what follows. The ground and excited energy levels measured from the graphene charge neutrality point read $E_g^{\rm Mo} = 27$ meV, $E_e^{\rm Mo} = 77$ meV, $E_g^{\rm W} = 230$ meV, and $E_e^{\rm W} = 398$ meV.
\begin{figure}
    \includegraphics[width=0.6\linewidth]{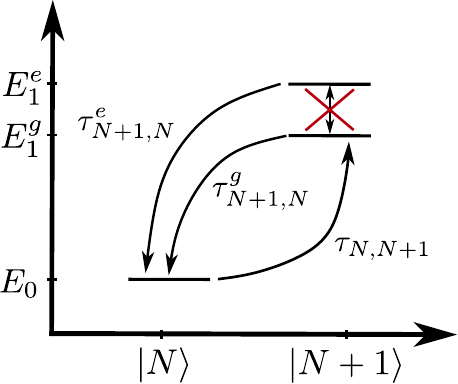}      
    \caption{Representation of the many-body Hilbert space and electron transition times, $\tau_{\alpha, \beta}$, between many-body states (labeled by particle number $N,N+1$) in the presence of spin-orbit coupling. 
    The energies are given in the grand canonical ensemble, $E_0 = \varepsilon_0 -N \mu$ and $E^{e,g}_1 = \varepsilon^{e,g}_1 -(N+1)\mu$, with $\mu \equiv E_\textnormal{F}$ at zero gate voltage and $\varepsilon_{0,1}$  are the total energies of the system with $N$ and $N+1$ electrons. Transitions that do not change the particle number are forbidden.
    }\label{f13} 
\end{figure} 

The SOC has no influence on the tunnelling Hamiltonian. Thus, starting from the model introduced in Sec. \ref{sec:model} and parametrized in Sec. \ref{sec:weak}, we construct the single-particle rate matrix from the functional form employed there and rotate it into the eigenstate basis with SOC. The single-particle tunneling rate is still diagonal in spin, but it mixes states with different angular momentum. 

A further simplification applies, though, if we additionally consider the secular or rotating wave approximation \cite{Darau2009}, which applies since $\hbar \Gamma \ll \Delta:= E_e  - E_g$. This approximation amounts to neglect the rapidly oscillating coherences, between states with different energies, as they average to zero on the typical time scales of the population dynamics. Eventually, the single-particle tunneling rate matrix can be taken completely diagonal. In the basis of the vacancy eigenstates, it reads $\bm{\Gamma}^{\rm SOC}(E) = \Gamma^{\rm SOC}(E) \otimes \bm{1}_4$
where 
\begin{equation}
\label{eq:G_SOC}
\Gamma^{\rm SOC}(E) = \dfrac{\Gamma_u(E) + \Gamma_{\bar{u}}(E)}{2},
\end{equation}
is the effective, still energy dependent, tunnelling rate. A direct extraction of the tunnelling amplitudes from the DFT band structures is, in this case, not possible, due to the interplay between the tunnelling coupling and the spin-orbit interaction, in determining the energy eigenstates.

\subparagraph{Transition rates.}
Employing Eqs. \eqref{e7} (and an analogous equation for the discharging rate) as well as $\bm{\Gamma}^\textnormal{SOC}$, we can compute the rates for the many-body transitions shown in Fig. \ref{f13}. Charging the vacancy occurs by tunneling of the electron from the graphene layer to one of the two quasi-degenerated levels of the vacancy, with transition time $\tau_{N, N+1}$. The transition rate reads
\begin{equation}\label{eq:W_SOC}
    W = \sum_{\ell\sigma} \Gamma^{\rm SOC}(\Delta E_{0, \ell\sigma})  f^+(\Delta E_{0, \ell\sigma}),
\end{equation}
which can be understood as $W = W_{0g} + W_{0e} $, as the final single-particle state in the many-body transition $N \rightarrow N+1$ is unknown. Each transition rate composing the total charging rate is given by
\begin{subequations}
\begin{align}
    W_{0g} &= 2\,\Gamma^{\rm SOC}(\Delta E_{0, g})f^+(\Delta E_{0, g}), \\ 
    W_{0e} &= 2\,\Gamma^{\rm SOC}(\Delta E_{0, e})f^+(\Delta E_{0, e}).
\end{align}
\end{subequations}
Discharging occurs by  tunneling of an electron originally in one of the two spin-orbit split Kramers' pairs towards the graphene bath. The corresponding transition rates are
\begin{subequations}\label{eq:Wg0We0}
\begin{align}
  W_{g0} &= 2\,\Gamma^{\rm SOC}(\Delta E_{0, g})f^-(\Delta E_{0, g}), \\ 
    W_{e0} &= 2\,\Gamma^{\rm SOC}(\Delta E_{0, e})f^-(\Delta E_{0, e}),
\end{align}
\end{subequations}
with transition times $\tau^g_{N+1, N}{:=} 1/W_{g0}$, $\tau^e_{N+1, N}{:=}1/W_{e0}$.

\begin{figure}
    \includegraphics[width=0.95\linewidth]{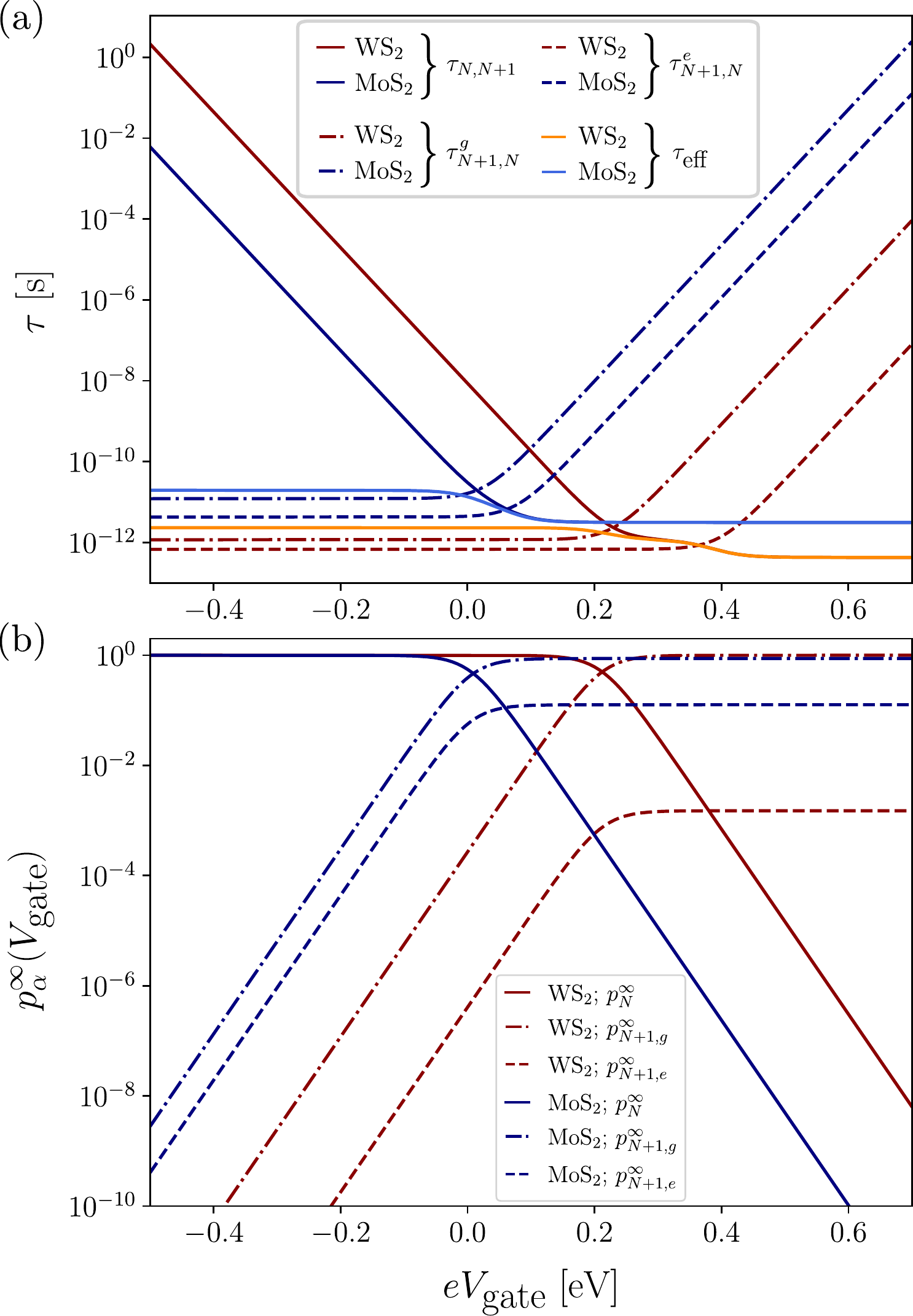}
    \caption{(a) Electronic transition times, $\tau$, for MoS\textsubscript{2} and WS\textsubscript{2} vacancy states as a function of the gate voltage, $V_\textnormal{gate}$ measured in eV. The solid, dashed and dashed-dotted lines correspond to the transitions $\alpha \rightarrow \beta$ shown in Fig. \ref{f13}.
    The dotted line corresponds to the effective transition time associated to the semiclassical dynamics defined from Eq. \eqref{eq:effectiverate_SOC} and controlling the dynamics from a local to global equilibrium state.
    For simplicity, we set the chemical potential at zero gate voltage to be zero. 
    (b) Corresponding stationary semiclassical probability of occupation of the many-body levels, $\mathbf{p}^\infty$, as a function of the gate voltage.
    }\label{f14} 
\end{figure}

We represent in Fig. \ref{f14} (a) these electron transition times as a function of the gate voltage, $V_\textnormal{gate}$. Compared
to the case where the spin-orbit coupling is negligible, we find a similar qualitative behavior of the transition times as a function of $V_\textnormal{gate}$. 
The charging rate is an effective rate were both transitions to the $g$ and $e$ states are allowed, thus the corresponding transition time shows a small jump when the chemical potential coincides with the higher energy level. This jump is better observed for WS\textsubscript{2} because the spin-orbit splitting is substantially larger than the characteristic thermal energy. In addition, at $V_\textnormal{gate} = 0$, where we measure the impact of spin-orbit interaction only, we find that comparing with the case where spin-orbit coupling is absent the charging process occurs faster for the WS\textsubscript{2} by almost one order of magnitude while being almost unchanged for MoS\textsubscript{2}, thus pointing to the spin-orbit interaction as a source of change of in the many-body transition rates.  

\subparagraph{Semiclassical dynamics.}

Using the transition rates, we derive a set of semiclassical dynamical equations for the populations using Eq. \eqref{eq:populations}. 
Differently from the case without spin-orbit interaction, the $\bar{\Gamma}$ point degeneracy splitting of the defect states, hinders the Lamb shift induced  dynamics. Technically, this happens since the Lamb shift Hamiltonian commutes with the reduced density matrix, $[\hat{H}_\textnormal{LS}, \hat{\rho}_\textnormal{red}] = 0$.  

For the calculation of the semiclassical dynamics, we assume, similarly to Sec. \ref{sec:weak}, local thermal equilibrium and, therefore, an equal  population of any degenerate level: $p_g  = p_{\bar{g}}$ and $p_e  = p_{\bar{e}}$.
The resulting dynamical equations can be written in matrix form as
\begin{equation}\label{eq:system_SOC}
   \dot{\mathbf{p}} = 
   \begin{pmatrix}
   -W & \tilde{W}_{g0} & \tilde{W}_{e0} \\
   W_{0g} & -\tilde{W}_{g0} & 0 \\
  W_{0e} & 0 & -\tilde{W}_{e0}
    \end{pmatrix}
    \mathbf{p},
\end{equation}
where $\mathbf{p} = (p_N \,\, p_{N+1,g} \,\, p_{N+1,e})^T$ and we remind that $\tilde{W}_\alpha := W_\alpha/2$. 
The local thermal equilibrium further imposes  a relation between  the populations of the single-particle states $|g \rangle$ and $|e \rangle$, namely, $p_e = p_g \exp (-\beta \Delta)$, where $\Delta = E_e - E_g$. Consequently, we  can reduce Eq. \eqref{eq:system_SOC} to a two-dimensional system of equations
\begin{equation}
 \begin{pmatrix}
  \dot{p}_N \\
  \dot{p}_{N+1}
 \end{pmatrix}
=
\begin{bmatrix}
 -W & f^-_0(\Delta) \tilde{W}_{g0} + f^+_0(\Delta) \tilde{W}_{e0} \\
  W & -f^-_0(\Delta) \tilde{W}_{g0} - f^+_0(\Delta) \tilde{W}_{e0}
\end{bmatrix}
\begin{pmatrix}
 p_N \\
 p_{N+1}
\end{pmatrix},
\end{equation}
where 
\begin{equation}
 f^\pm_0(\Delta) := \dfrac{1}{1 + \exp(\pm \beta \Delta)}.
\end{equation}
These equations are solved under the constraint $\sum_\alpha p_\alpha = p_N + p_{N+1} = 1$ with  $p_{N+1} = p_g + p_e$. The analytical solution of the system of equations yields 
\begin{equation}
 \mathbf{p}(t) = \mathbf{p}^\infty + p_0 
 \begin{pmatrix}
  1 \\
  -f^-_0(\Delta) \\
  -f^+_0(\Delta) 
 \end{pmatrix}
{\rm e}^{-W_\textnormal{eff} t},
\end{equation}
where we employed the inverse relations between $p_{N+1}$ and the populations of the $g$ and $e$ levels, $p_g = f_0^-(\Delta) p_{N+1}$ and $p_e = f^+_0(\Delta) p_{N+1}$, as well as defined the effective decay rate
\begin{align}
 W_\textnormal{eff} &= W + f^-_0(\Delta) \tilde{W}_{g0} + f^+_0(\Delta) \tilde{W}_{e0}, \notag
 \\ & = \Gamma^{\textnormal{SOC}}(\Delta E_{0,g})[2 f^+(\Delta E_{0,g}) + f_0^-(\Delta)f^-(\Delta E_{0,g})] \notag \\
 &+ \Gamma^{\textnormal{SOC}}(\Delta E_{0,e})[2 f^+(\Delta E_{0,e}) + f_0^+(\Delta)f^-(\Delta E_{0,e})] \label{eq:effectiverate_SOC}
\end{align}
These expressions reduce to Eqs. \eqref{eq:rates_t}-\eqref{eq:rate_eff} for the limit $\Delta \rightarrow 0$ in which all four states of the one particle sector are degenerated.

Eq. \eqref{eq:effectiverate_SOC} defines the effective relaxation time $\tau_\textnormal{eff}: = 1/W_\textnormal{eff}$ for the population dynamics in the presence of spin-orbit coupling. We show $\tau_\textnormal{eff}$ for our heterobilayers in comparison to $\tau_{N, N+1}$ and $\tau_{N+1, N}$ in Fig. \ref{f14} (a). As in the case of Fig. \ref{f10} (a), this effective time is less sensitive to the gate voltage compared to $\tau_{N, N+1}$ and $\tau_{N+1, N}$. More importantly, when looking at $\tau_\textnormal{eff}$ for WS\textsubscript{2}--Gr at $V_\textnormal{gate} = 0$, we find $\tau^{\rm W}_\textnormal{eff} \simeq 2.3$ ps at $\mu =0$ (for comparison, $\tau^{\rm Mo}_\textnormal{eff} \simeq 13.5$ ps for the MoS\textsubscript{2}-Gr interface). Shift of the vacancy levels following the G\textsubscript{0}W\textsubscript{0} for the pristine system, can reduce the relaxation time by at most $\sim 25\%$. In any case, this value is of the order of magnitude of the typical electronic relaxation times measured by tr-ARPES for the same type of heterostructure and same process \cite{Gierz2020, Gierz2021}. 
Note that within our model, we can account for the effect on the relaxation times of changes of the chemical potential due to doping at the graphene layer; our calculated relaxation times can change up to one order of magnitude within the range from $-300$ to $400$ meV.
Our results therefore suggest that, if the vacancy concentration is low, defect charge transfer can be a potential mechanism that participates in the ultrafast relaxation times observed for photoexcited electrons in these heterobilayers.

\begin{figure}
    \includegraphics[width=1.0\linewidth]{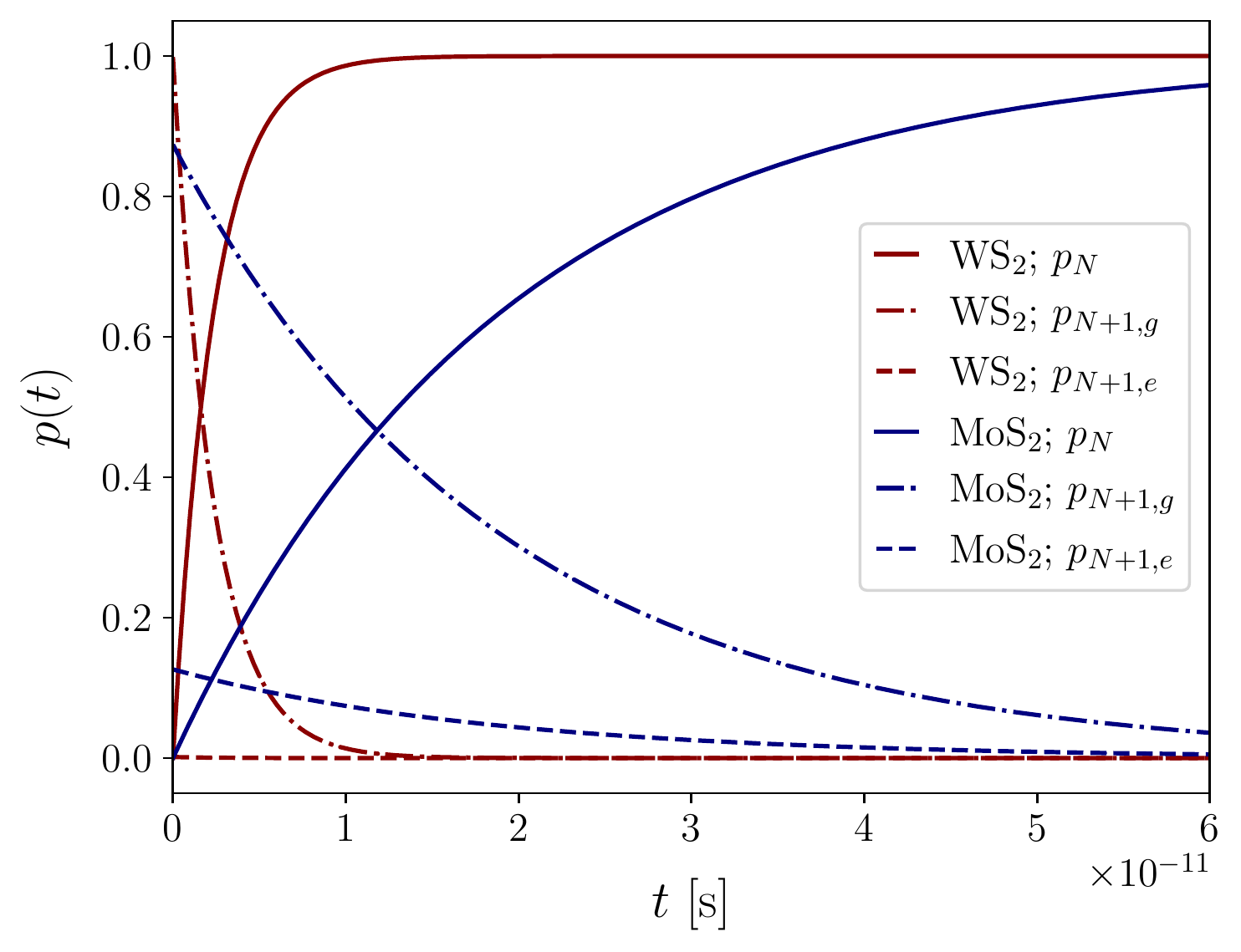}
    \caption{Population, $p(t)$, of the energy levels as a function of time, $t$ (measured in s). The dynamics is computed for the representative gate voltage, $V_\textnormal{gate} = -0.3$ eV.
    }\label{f17} 
\end{figure}

The vector $\mathbf{p}^\infty$ yields the stationary population at a given chemical potential and temperature of the bath for the states $|N \rangle$, $|g \rangle$ and $|e \rangle$,
\begin{equation}\label{eq:populations_SOC}
 \mathbf{p}^\infty = \dfrac{W}{W_\textnormal{eff}}
 \begin{pmatrix}
  W_\textnormal{eff}/W - 1 \\
  f^-_0(\Delta) \\
  f^+_0(\Delta) 
 \end{pmatrix}.
\end{equation}
In Fig. \ref{f14} (b), we display the stationary population \eqref{eq:populations_SOC} for the two studied interfaces and evaluated for the same range of gate voltages and physical parameters as in Fig. \ref{f14} (a). The values of the stationary population are controlled by the many-body transition rates, as well as temperature. In the large negative $V_\textnormal{gate}$ limit, the chemical potential is far below the available defect levels and therefore, the vacancy remains empty, $p_N \rightarrow 1$ and $p_g, p_e \rightarrow 0$. After increasing $V_\textnormal{gate}$ and once the chemical potential is larger than the single-particle energy of the empty defect states, tunneling becomes possible and the vacancy charges at equilibrium, with the relation between the $g$ and $e$ levels dictated by the Boltzmann factor, $\exp(-\beta \Delta)$. Since the splitting of the defect levels due to spin-orbit interaction is substantially smaller in MoS\textsubscript{2} compared to WS\textsubscript{2}, the population of $|e \rangle_\textnormal{Mo}$ remains several orders of magnitude larger in the large positive  $V_\textnormal{gate}$ limit compared to $|e \rangle_\textnormal{W}$.
Finally, as in Sec. \ref{sec:weak}, the value for $p_0$ depends on the initial conditions for the time evolution. It can be easily calculated, for example, $p_0 = 1-W/W_\textnormal{eff}$ if the initial condition is fixed to $p_N = 0$ (vacancy is occupied and therefore the system decharges over time).
As an example, we show in Fig. \ref{f17} charge quenching as given by Eq. \eqref{eq:populations_SOC} for fixed gate energy $e V_\textnormal{gate} = -0.3$ eV. The dynamics shows the expected exponential decay law, controlled by $W_\textnormal{eff}$ as well as the asymptotic behavior seen in Fig. \ref{f14} (b). Also, as anticipated, if the temperature energy scale $\beta$ is smaller than the splitting $\Delta$, thermal population of the $|e\rangle$ state does not occur, as for WS\textsubscript{2} compared to MoS\textsubscript{2}. 

\section{Conclusion}\label{sec:conclusion}

In this work, we have investigated, by a combination of density functional theory, model Hamiltonian and master equation techniques, how defect states created by monoatomic chalcogen vacancies in representative MoS\textsubscript{2}--Gr and WS\textsubscript{2}--Gr interfaces are an interesting playground for electronic dynamical processes.
From the \textit{ab initio} calculations, we find that the empty vacancy levels are naturally hybridized with the Dirac cone 
from the graphene layer.
We propose a low-energy model and derive the single-particle tunneling rate matrices employing information from the DFT band structure calculation as well as group theory arguments. This allows us to gain physical insight into the impact of symmetries into the parameter-free many-body transition rates.
At the level of sequential tunneling, we perturbatively computed the transition rates for the population and depopulation of the vacancy by one electron tunneling to or from the graphene reservoir.  
We find that transition rates (and corresponding electronic transition times) are sensitive to external electric fields perpendicular to the heterobilayer plane, and created by a voltage gate on graphene, changing also by several orders of magnitude.
A shift of the chemical potential via this gate voltage can also change the relative order of the rates for the different TMDC (MoS\textsubscript{2}, WS\textsubscript{2}) considered in this work. This is shown to influence the semiclassical population dynamics, derived here within an effective two-level model assuming local thermal equilibrium.
For strong spin-orbit interaction, we proposed a modified transport model for the charge transfer using the DFT band structure information as well. We obtain a single-particle rate matrix which is still proportional to the identity, but with non-trivial energy dependence.
We computed the corresponding transition rates at the level of sequential tunneling and employing the secular approximation. A three-level model for the dynamics of the population and depopulation of the vacancy has been proposed and solved assuming local thermal equilibrium.  The effect of spin-orbit interaction has been gauged in our charge transfer rates. 

Our results for the effective rate suggests that defects slow down the charge transfer, effectively trapping electrons, and can be a relevant transport channel in the ultrafast charge transfer processes observed in recent experiments \cite{Gierz2020, Gierz2021} assuming the vacancy concentration considered in this work. 
Our research serves as a basis to study other more subtle dynamical features associated to quantum internal dynamics occurring at short time scales due to quantum degeneracy of the vacancy energy levels or electron-phonon coupling \cite{Zhen2017}, and we demonstrated how simple dynamical models can be used as guidance in the understanding of experimental transport phenomena at these heterostructures and complex interfaces.


\begin{acknowledgments}
Discussions with F. Evers, I. Gierz, M. Glazov, E. Malic, R. Perea-Caus\'in, D. Qiu, A. Rappe and H. Steinberg are gratefully acknowledged. We are thankful to S. Latini, L. Xian and A. Rubio for providing us with the non-defected geometry of the heterostructure employed in some of the calculations in this article.
The computations were performed using the resources of the Chemfarm local cluster at the Weizmann Institute of Science and the MPCDF computing center. D. H.-P. and S. R. A. acknowledge funding from a Minerva Foundation grant. D. H.-P., A. D. and S. R. A. acknowledge funding from the Collaborative Research Center SFB 1277 (Project-ID 314695032, subprojects B02 and B10). 

\end{acknowledgments}

\bibliography{biblio}

\end{document}


\title{{Supplemental Material \\ Charge quenching at defect states in  transition metal \\ dichalcogenide--graphene  van der Waals heterobilayers}}

\author{Daniel \surname{Hernang\'{o}mez-P\'{e}rez}}


\affiliation{Department of Molecular Chemistry and Materials Science, Weizmann Institute of Science, Rehovot 7610001, Israel}

\author{Andrea Donarini}

\affiliation{Institute of Theoretical Physics, University of Regensburg, 93040 Regensburg, Germany \vspace{0.3cm} \\ \textnormal{Email: daniel.hernangomez@weizmann.ac.il}}

\author{Sivan Refaely-Abramson}


\affiliation{Department of Molecular Chemistry and Materials Science, Weizmann Institute of Science, Rehovot 7610001, Israel}

\keywords{2D materials, transition-metal dichalcogenides, van der Waals, defects, graphene, transport, single-electron}

\maketitle


\tableofcontents
\clearpage

\section{Computational details}\label{app:computational}
Our \textit{ab initio} calculations were performed using  
density functional theory as implemented in the 
\textsc{quantum espresso} package \cite{Giannozzi2009}.
For the exchange-correlation functional, we used
the non-empirical generalized gradient approximation of Perdew-Burke-Ernzerhof  \cite{Perdew1996}. 
We employed a plane-wave basis set and scalar relativistic (full relativistic for the spin-orbit coupling calculations) 
norm-conserving pseudopotentials \cite{Dojo2019} with
a basis cut-off of $95$ Ry for WS\textsubscript{2}--Gr and $90$ Ry for MoS\textsubscript{2}--Gr.
The self-consistent charge density was calculated within a $6 \times 6 \times 1$ grid
and the total energy considered to be converged when the difference between iterations was
smaller than a threshold value of $10^{-9}$ Ry.
Preoptimizations of the supercell were performed with VASP \cite{Kresse1996}, using LDA as exchange correlation functional, a basis energy cut-off of $600$ eV and a $6\times6\times1$ \textbf{k}-grid.
Additional relaxations in the presence of the vacancy were performed by fixing the supercell lattice vectors and optimizing the position of the atoms during the self-consistent field cycle until all the components of the ionic forces were smaller than $10^{-3}$ Ry/$a$\textsubscript{0}.
For the optimization procedure we also employed the van-der Waals corrected functional \texttt{vdw-df-09} \cite{Troy2009, Berland2015} to account for potential additional changes in the interlayer separation.

\section{Elements of applied group theory}\label{app:symmetry}

\begin{figure}
    \includegraphics[width=0.45\linewidth]{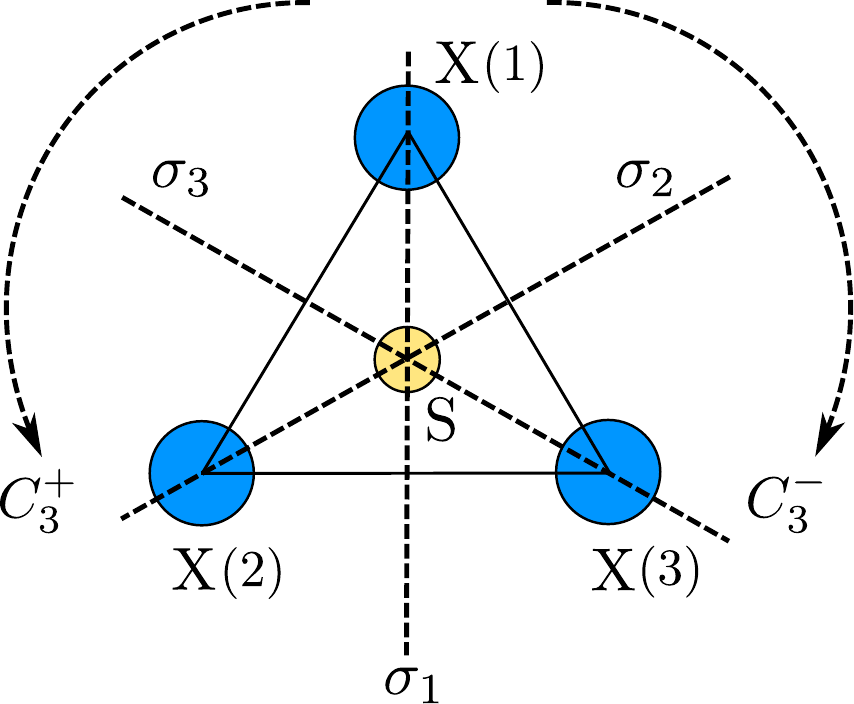}
    \caption{Schematic representation of the non-trivial operations of the group C\textsubscript{3v} (rotation by $2 \pi/3$, reflection with respect to a plane of symmetry) for the chalcogen vacancy in the XS\textsubscript{2} monolayer.
    }\label{f9} 
\end{figure}

For the sake of completeness, we briefly summarize in section of the Supporting Information (SI) elementary point-group theory results that underlay the density functional theory findings and the calculation of the single-particle tunneling rate matrix in the main paper and Sec. \ref{app:coefficients_nosoc} of the SI. We 
follow here Refs. \onlinecite{Wigner1959, Dresselhaus2008, Cotton1990}.
In the absence of the graphene layer, the XS\textsubscript{2} TMDC layer with a chalcogen vacancy possesses
C\textsubscript{3v} symmetry, the point-group of the equilateral triangle.
This point group is characterized by a proper rotation axis of order $3$ and 
 $3$ planes of symmetry. Traditionally, E, $C_3^{\pm}$ and $\sigma_v$ ($v = 1,2,3$)
are chosen as notation for the identity, rotation and reflection elements of the group, defining  
three equivalence classes (see scheme in Fig. \ref{f9}).
%
%

As a consequence of Schur's lemmas \cite{Dresselhaus2008}, the eigenvalues and eigenstates of a Hamiltonian which commutes with all symmetry operators $\hat{G}$ of a group can be classified according to the irreducible represetations of the group itsself. This is the case, if and only if, the group is a complete group of invariance.
In other words, we can use the unique reduction of the group into its irreducible representations to find a block-diagonal form of the Hamiltonian, with its invariant subspaces having the same dimension as the irreducible representations of the group. 
Moreover, the number of irreducible representations is equal to the number of equivalence classes in the symmetry group. In the particular case of C\textsubscript{3v}, the group has three non-trivial irreducible representations of dimension $1$ and $2$ [typically noted
as $A_1, A_2, E$, with $\textnormal{dim}(A_i) = 1$ and $\textnormal{dim}(E) = 2$].
Consequently, if the defect preserves the original symmetry of the lattice, the defect states are at most double degenerated.
In the case of the XS\textsubscript{2}-defected monolayers considered in this work, three defect states appear close to the TMDC: a pair of empty in-gap degenerated E states and an occupied non-degenerated A\textsubscript{1} state close to the topmost valence band. Note that the energetics of the defect states cannot be predicted by group theory only, therefore, pointing towards the need of employing \textit{ab initio} methods in the study of these systems.

Van der Waals adsorption of graphene to the TMDC layer  
lowers the global symmetry of the supercell from C\textsubscript{3v} to C\textsubscript{s}.
The C\textsubscript{s} symmetry group only has one element apart from the identity $E$, given by the plane of symmetry and noted $\sigma_s$.
As a difference to C\textsubscript{3v}, C\textsubscript{s} is Abelian (all the elements of the group commute), and all its irreducible representations are forcibly one-dimensional. 
Consequently, the eigenenergies and eigenstates of the system are not guaranteed to be degenerated (except for accidental degeneracies). 
The heterobilayer eigenstates stemming from the former vacancy E states, can now be classified according to one the reflection operators, $\sigma_s$. Their different nature is associated to the different size  of the anti-crossings between the impurity band and the Dirac cone of graphene. We will use this observation in the calculation of the single-particle rate matrix in Sec. \ref{app:coefficients_nosoc} of the SI.

\section{Calculation details for $\Gamma^p_{ij}(E,\sigma)$}\label{app:coefficients_nosoc}
In this section, we derive an expression for the single-particle rate matrix $\Gamma^p_{ij}(E,\sigma)$ based on the group theory analysis
from Sec. \ref{app:symmetry} of the SI. We start from Eq. (19), which contains as crucial component the tunneling (``hybridization'') matrix elements, $t_{i\mathbf{k}\sigma}$. These matrix elements are assumed to be spin-diagonal, $t_{i\sigma,\mathbf{k}\sigma'} = t_{i\mathbf{k}\sigma} \delta_{\sigma\sigma'}$, as the tunneling occurs over a thin vacuum layer which cannot change the spin orientation.
By definition, $ t_{i\mathbf{k}\sigma}$ are given by
 \begin{equation}\label{e1}
     t_{i\mathbf{k}\sigma} = \langle i \sigma | \hat{H} | \mathbf{k} \sigma\rangle := \langle i | \hat{H} | \mathbf{k} \rangle,
 \end{equation}
where $\hat{H}$ is the Hamiltonian of the system, $|i\rangle \equiv |i,\mathbf{0} \rangle$ is a localized impurity state, localized, for convenience, at the origin of the Bravais supercell lattice,  $|\mathbf{k} \rangle$ a graphene eigenstate and the spin degree of freedom is kept implicit as it does not intervene in the calculation.
 
We can express the defect band in the reciprocal space as
\begin{equation}\label{eq:bloch_impurity}
 |i,\mathbf{k} \rangle = \dfrac{1}{\sqrt{N_\textnormal{sc}}} \sum_{\mathbf{R}} {\rm e}^{\mathfrak{i} \mathbf{k} \cdot \mathbf{R}} |i,\mathbf{R} \rangle,
\end{equation}
where $\{\mathbf{R}\}$ spans the supercell lattice and $N_\textnormal{sc}$ is the number of supercell lattice points.
It is convenient to have  the  defect at the origin of coordinates (see Fig. 1 in the main text); then, the inverse transformation reads
\begin{equation}
|i,\mathbf{0} \rangle = \dfrac{1}{\sqrt{N_\textnormal{sc}}} \sum_\mathbf{k} |i,\mathbf{k} \rangle,
\end{equation}
where $\{\mathbf{k}\}$ spans the quasi-momenta lattice of the supercell Brillouin zone.

Using Eq. (19), we estimate the value of the single-particle rate matrix by looking at the region of the spectrum showing level repulsion between the graphene and the vacancy states. Note that we consider only the orbital component of the system without spin-orbit interaction, spin-orbit coupling will be incorporated perturbatively into the model at a later stage in the main paper. Substituting the matrix elements we find, for each spin sector,
\begin{align}
    \Gamma^+_{ij}(E) &= \dfrac{2\pi}{\hbar} \dfrac{1}{N_\textnormal{sc}} \sum_{\mathbf{k}} \sum_{\mathbf{k}',\mathbf{k}'' } 
    \langle i,\mathbf{k}'| \hat{H}| \mathbf{k} \rangle \langle  \mathbf{k} | \hat{H}| j,\mathbf{k}'' \rangle 
    \delta(E- \epsilon_{\mathbf{k}}),  \\
    & = \dfrac{1}{N_\textnormal{sc}} \sum_{\mathbf{k}} 
    \langle i,\mathbf{k}| \hat{H}| \mathbf{k} \rangle \langle  \mathbf{k} | \hat{H}| j,\mathbf{k} \rangle \delta(E- \epsilon_{\mathbf{k}}),\label{eq:gamma_ij}
\end{align}
where we used that the graphene and the TMDC layers have the same periodicity. In other words, the translational invariance with respect to the supercell lattice vectors ensures that $t_{i\mathbf{k}\sigma}$ only couples graphene and vacancy states with the same quasi-momentum.

The single-particle tunneling rate matrix is not diagonal in the quasi-angular momentum basis labeled by $i,j = \pm$ because the graphene substrate does not share the same symmetry as the TMDC layer. Instead, as discussed in Sec. \ref{app:symmetry} of the SI, in the presence of the graphene layer the TMDC global C\textsubscript{3v} symmetry is lowered to a C\textsubscript{s} symmetry. 
Consequently, we can classify the states by using the action of an operator from  the group C\textsubscript{s}. For example, we can 
choose the major plane of symmetry on Fig. 1 whose action on Eq. \eqref{eq:bloch_impurity} is
\begin{equation}
\hat{\sigma}_s |+,\mathbf{k} \rangle = \dfrac{1}{\sqrt{N_\textnormal{sc}}} \sum_\mathbf{R} {\rm e}^{\mathfrak{i} \mathbf{k} \cdot \mathbf{R}} |-, \sigma_s \mathbf{R} \rangle = |-, \sigma_s\mathbf{k} \rangle.
\end{equation}
Similarly, we find $\hat{\sigma}_s |-,\mathbf{k} \rangle =  |+ ,\sigma_s\mathbf{k} \rangle$. 
We therefore construct a basis from the quasi-angular momentum states in which $\hat{\sigma}_s$ is diagonal, $\hat{\sigma}_s |u \rangle = u |u \rangle$, with $u = \pm$ from the linear combination (normalized sum/difference) of the $i=\pm$ states. In this basis,
$\hat{\sigma}_s |u,\mathbf{k} \rangle = u |u,\sigma_s \mathbf{k} \rangle $ and $ \hat{\sigma}_s |\mathbf{k} \rangle = \exp[i \varphi(\mathbf{k})] |\sigma_s \mathbf{k} \rangle $ where $\varphi(\mathbf{k})$ is a phase.
We now show that Eq. \eqref{eq:gamma_ij} is a diagonal matrix in this basis, which also defines a symmetry of the Hamiltonian $\hat{H}$. 
Using the observations above, it is straightforward to prove that
\begin{equation}
\sum_{\mathbf{k}} \langle u,\mathbf{k}  | \hat{H} | \mathbf{k} \rangle \langle \mathbf{k} |\hat{H} | \bar{u} \mathbf{k}  \rangle =
\sum_{\mathbf{k}} \langle u,\mathbf{k}  | \hat{\sigma}_s^2\hat{H}  \hat{\sigma}_s^2 |\mathbf{k} \rangle 
                  \langle \mathbf{k} |\hat{\sigma}_s^2\hat{H} \hat{\sigma}_s^2| \bar{u} \mathbf{k}  \rangle =
- \sum_{\mathbf{k}'} \langle u,\mathbf{k}' | \hat{H} | \mathbf{k}' \rangle \langle \mathbf{k}' |\hat{H} | \bar{u},\mathbf{k}' \rangle,
\end{equation}
with $\mathbf{k}'  = \sigma_s \mathbf{k}$ and $\bar{u} = -u$. Here, we used that the phase $\varphi(\mathbf{k})$ cancels out into the projector $|\mathbf{k} \rangle \langle \mathbf{k}|$. 
As a consequence $\Gamma^+_{u\bar{u}} = - \Gamma^+_{u{\bar u}} = 0$, i.e. the single-particle rate matrix is diagonal in the orbital subspace, $\Gamma_{uu'} = \Gamma^+_{uu} \delta_{uu'}$ with $\Gamma^+_{uu'} = \Gamma^{-}_{u'u}$.

We thus consider only the diagonal elements,
\begin{equation}
 \Gamma_{uu}^+(E) = \dfrac{2 \pi}{\hbar} \dfrac{1}{N_\textnormal{sc}} \sum_\mathbf{k} |\langle u,\mathbf{k} | \hat{H}| \mathbf{k} \rangle|^2 \delta(E- \epsilon_{\mathbf{k}}),
 \label{eq:gamma_uu}
\end{equation}
and, in the limit $N_{sc} \to \infty$, we transform the discrete sum over the quasimomenta into an integral over the supercell Brillouin zone by
\begin{equation}
    \sum_\mathbf{k} \rightarrow \dfrac{N_\textnormal{sc}  A_\textnormal{sc}}{(2\pi)^2} \int_{\rm SBZ} \!\!\!\!\!\textnormal{d}\mathbf{k},
\end{equation}
%
where $A_{sc}$ is the supercell area. Thus, Eq. \eqref{eq:gamma_uu} becomes
\begin{equation}
    \Gamma_{uu}(E) = \dfrac{2 \pi}{\hbar} \dfrac{A_\textnormal{sc}}{(2 \pi)^2} |t_u|^2 \sum_\tau \int_0^{2\pi} d\varphi \int_0^{\infty} dk k \delta (E - \hbar v_\textnormal{F} k),
    \end{equation}
where $\tau = \pm$ is the valley index, $v_\textnormal{F}$ the Fermi velocity and $k$ measures the distance of the quasi-momentum from a graphene Dirac point.
Moreover, $t_u :=t_{u\mathbf{k}\sigma} $ is defined as in Eq. \eqref{e1} and we dropped the $\pm$ superindex as the matrix is diagonal. Notice that the integral has been extended to $k \rightarrow +\infty$, which can be justified by the $\delta$ function, as far as $E$ is sufficiently close to the graphene charge neutrality point. 

Introducing the graphene unit cell, $A_\textnormal{c}$ and density of states close to its pristine chemical potential, $\rho_\textnormal{Gr}(E) = 2 A_c |E|/(\pi \hbar^2 v_\textnormal{F}^2)$, we obtain the diagonal elements of the single-particle rate matrix in their final form
\begin{equation}
 \Gamma_{uu}(E) = \dfrac{\pi^2}{2\hbar}  \dfrac{A_\textnormal{sc}}{A_\textnormal{c}} |t_u|^2 \rho_\textnormal{Gr}(E).
\end{equation}

The final step consists in estimating the absolute value of the hybridization matrix elements, which can be done by employing Eq. (23) in the main text in the context of avoided crossings in a local two-level system.

\section{Temperature and defect energy dependence of the many-body transition rates}\label{sec:temperature}

\begin{figure}
\vspace{0.0cm}
    \includegraphics[width=0.65\linewidth]{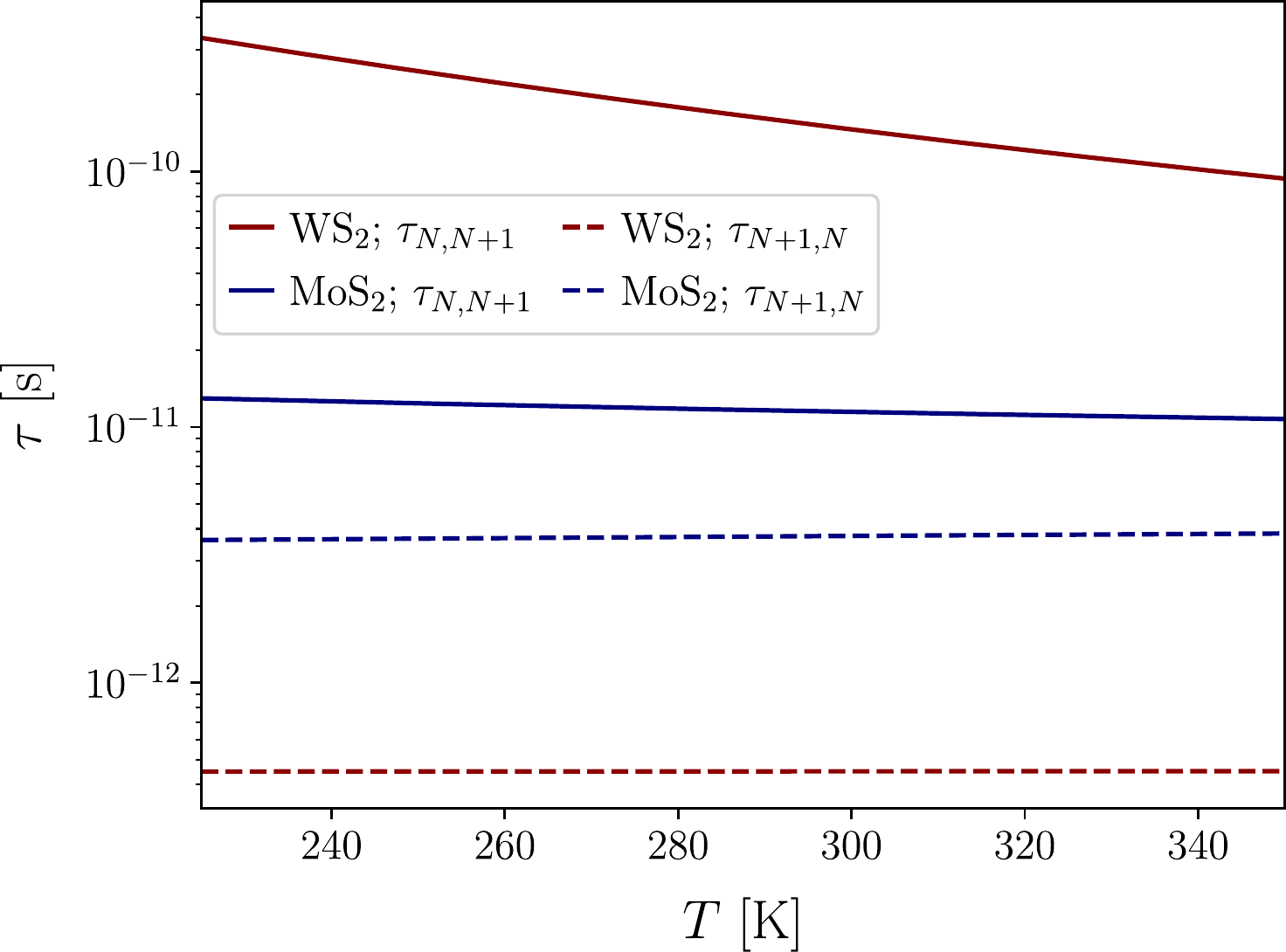}
    \caption{Electronic transition time, $\tau$, for MoS\textsubscript{2} and WS\textsubscript{2} vacancy states as a function of bath temperature, $T$ measured in kelvin, for the transitions $\alpha \rightarrow \beta$ with $\alpha, \beta \in \{N,N+1\}, \alpha \neq \beta$.
    }\label{f11} 
\end{figure}

We display in Fig. \ref{f11} the transition times evaluated for a wide range of bath temperatures at fixed chemical potential (here $\mu = 0$). For this temperature, well within the weak coupling limit $\hbar \Gamma \ll k_\textnormal{B} T$, the transition rates are very smooth functions. Temperature opens (or closes) the window around the defect energies where the transition to and from the reservoir can be active, and therefore, larger sensitivity to temperature is expected in WS\textsubscript{2} compared to MoS\textsubscript{2} because its defect levels are further from the Fermi energy of undoped graphene.
In addition, as we are essentially observing the effect of temperature on the Fermi function, similar behavior occurs for the case of strong spin-orbit interaction by evaluation of Eqs. (39)-(41).

\begin{figure}
\vspace{0.0cm}
    \includegraphics[width=0.65\linewidth]{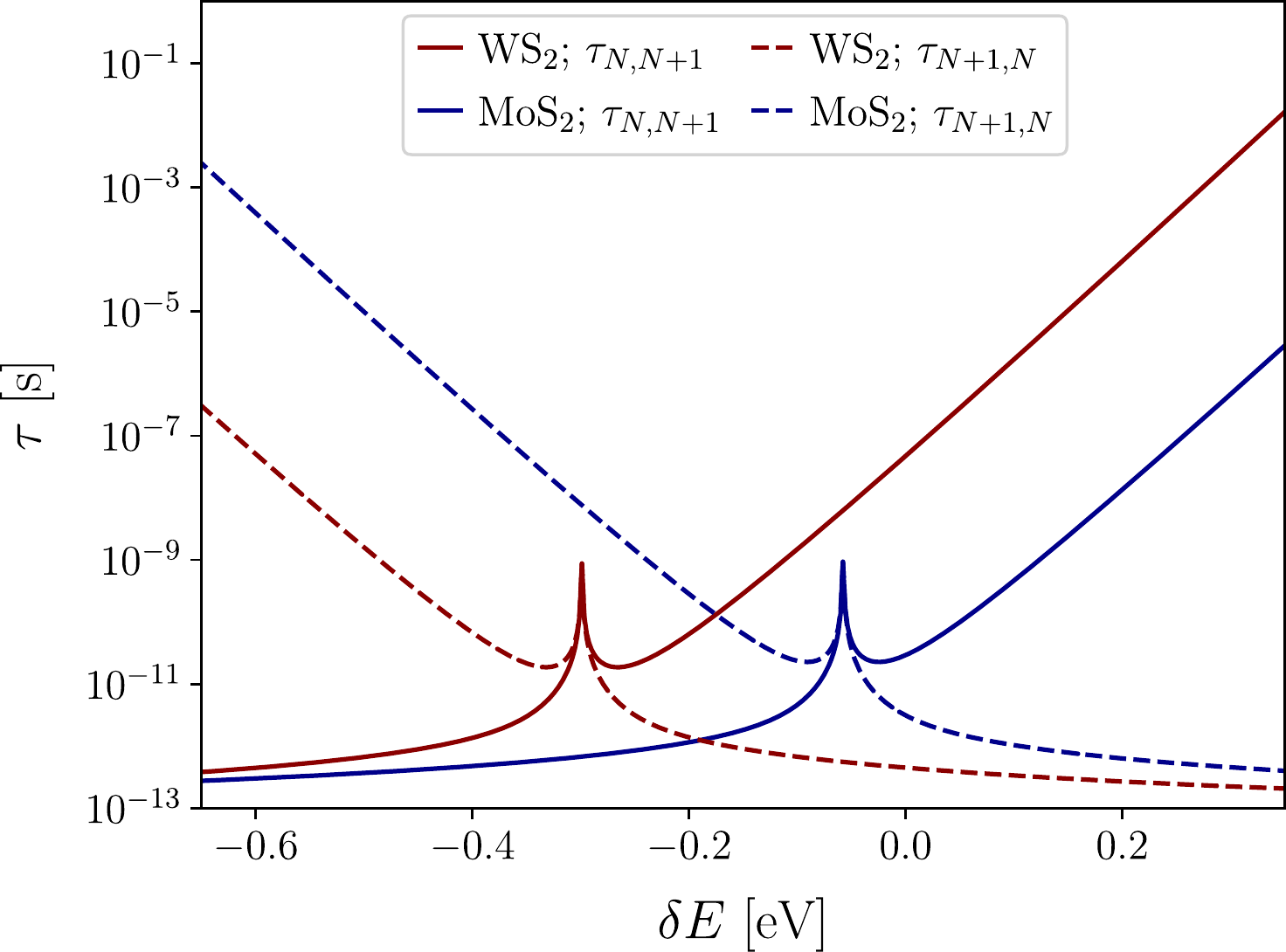}
    \caption{Electronic transition times, $\tau$, for MoS\textsubscript{2} and WS\textsubscript{2} vacancy states as function of the shift of the defect levels, $\delta E$, given in eV and measured from  $\mu = 0$. This shift would result from \textit{e.g.} including screening effects. The solid and dashed lines correspond to the transitions $N \rightarrow N +1$ and $N + 1 \rightarrow N$ respectively.
    }\label{f15} 
\end{figure}

\begin{figure}
\vspace{0.5cm}
    \includegraphics[width=0.65\linewidth]{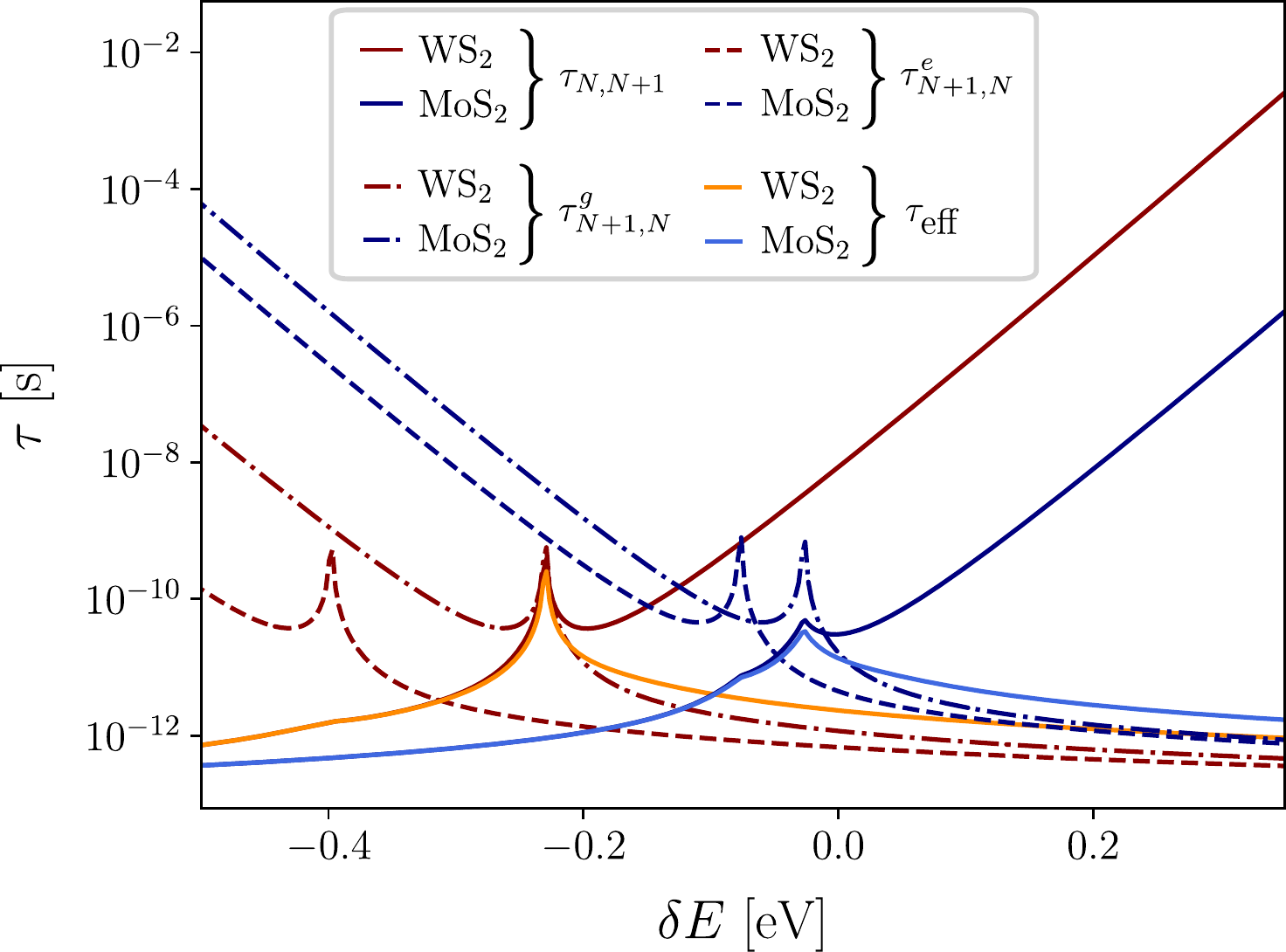}
    \caption{Electronic transition times with spin-orbit interaction, $\tau$, for vacancy states at the MoS\textsubscript{2}--Gr and WS\textsubscript{2}--Gr interface as function of the shift of the defect levels, $\delta E$, given in eV and measured from  $\mu = 0$. This shift would result from \textit{e.g.} including screening effects. The solid and dashed lines correspond to the transitions $N \rightarrow N +1$ and $N + 1 \rightarrow N$ respectively.
    }\label{f16} 
\end{figure}

In Figs. \ref{f15} and \ref{f16}, we show the change in the transition time when the position of the defect levels with respect to the charge neutrality point is rigidly shift. We assume that this shift is not too large so that the hybridization matrix elements remain constant. This type of level shift can be used to understand the impact of the change of the position of the defect levels in the transition rates at lowest order, if screening would be taken into consideration, \textit{e.g.} at the G\textsubscript{0}W\textsubscript{0} level. 
We consider first on the case of weak spin-orbit interaction. Overall, we observe that the trends corresponding to the charging and discharging processes are reversed when comparing to the trends in Fig. 5 (a). For example, the charging of the vacancy becomes slower when the energy levels are shifted away from the charge neutrality point, while the discharging becomes faster. In other words, the positive rigid shift that would occur if screening would be included works in the opposite way as the positive gate voltage (as the pristine band gap becomes larger). 
In addition, a cusp over the smooth trends is present once the defect levels are shifted towards the graphene Dirac point. This cusp is originated by the alignment of the defect energy with the energy at the $\bar{\textnormal{K}}$ point, $\delta E = -\Delta E_{0,u}$, where the inverse DoS acquires a  non-analytical scaling as $\sim |E|^{-1}$. While an enhancement of the relaxation time should be expected due to these phase space arguments, its divergence is  an artefact of the perturbative approach. For example, when considering the tunnelling coupling in the dressed second order approximation \cite{Kern2013}, the tunnelling rate for a given many-body transition is obtained by convolution of the DoS with a Lorentzian centered on the resonant energy, thus preventing  exact vanishing of the rate. In Fig.~\ref{fig16} we present the electronic transition times in presence of SOC. The transition times to the single energy levels are still characterized by the diverging cusp, which is smoothed in $\tau_{\rm eff}$, as the effective rate never vanishes due to the contribution of both energy levels.
Evidently, at the G\textsubscript{0}W\textsubscript{0} level the precise value of  $\delta E$ is fixed once the screening conditions are determined, but the numerical simulations suggest both an additional way of modifying the transition rates if the dielectric environment is altered in a controlled way and a manner of interpret and understand screening from the experimentally measured transition rates.

\bibliography{biblio.bib}